\documentclass{iopjournal}
\usepackage{amsmath}
\usepackage{amssymb}
\usepackage{bm}
\usepackage{dcolumn}
\usepackage{hyperref}
\usepackage{cite}

\begin{document}

\articletype{Paper}

\title{Engineering many-body tunneling dynamics of one-dimensional bosons with finite-core power-law interactions}

\author{
Abhik Kumar Saha$^{1}$\orcid{0000-0001-8168-6742},
Cesare Vianello$^{2,3}$\orcid{0009-0001-1136-5924},
Andrea Trombettoni$^{4,5}$\orcid{0000-0002-1108-4727}
and Barnali Chakrabarti$^{6}$\orcid{0000-0002-6320-9894}
}

\affil{$^1$Department of Physics, Kyoto University, Kitashirakawa Oiwakecho, Sakyo-ku, Kyoto 606-8502, Japan}

\affil{$^2$Dipartimento di Fisica e Astronomia ``Galileo Galilei'',Universit\`a di Padova, Via Marzolo 8, 35131 Padua, Italy}

\affil{$^3$INFN Sezione di Padova, Via Marzolo 8, 35131 Padua, Italy}

\affil{$^4$ Department of Physics, University of Trieste, Strada Costiera 11, 34151 Trieste, Italy.}

\affil{$^5$ CNR-IOM DEMOCRITOS Simulation Centre and SISSA, Via Bonomea 265, 34136 Trieste, Italy.
}

\affil{$^6$Laborat\'orio de F\'isica Te\'orica e Computacional,
Departamento de F\'isica, Universidade Federal de Pernambuco,
50670-901 Recife, Pernambuco, Brazil}

\email{saha.abhikkumar.3k@kyoto-u.ac.jp}
\email{barnali.chakrabarti@ufpe.br}

\begin{abstract}

We investigate the many-body dynamics of a one-dimensional Bose-Josephson junction realized by ultracold bosons in a box-like potential with a central barrier, interacting via two-body finite-core power-law interactions $V_\alpha(r)=g/(r^\alpha+d^\alpha)$, where $d$ sets the short-distance interaction scale and $\alpha$ controls the spatial dependence of the interaction. Using the multiconfigurational time-dependent Hartree method for bosons (MCTDHB), we  systematically explore how the interaction profile and strength, the central barrier, and the initial population imbalance shape the tunneling dynamics. At weak interaction, increasing $\alpha$ drives a crossover from usual Josephson oscillations to dephasing followed by recurrent rephasing, demonstrating that the dynamics is controlled by the full spatial structure of the interaction kernel. A higher barrier suppresses the recurrent dynamics, whereas a stronger initial population imbalance promotes relaxation towards a nearly vanishing population imbalance. Increasing the interaction strength progressively suppresses population-transfer fluctuations, with this suppression becoming particularly relevant at larger $\alpha$. A comparison with the two-mode extended Bose-Hubbard model shows that the latter captures the dynamics when the evolution is dominated by the lowest orbital in each well, but progressively loses accuracy as additional orbitals become dynamically populated. In the strongly interacting regime, many-body spatial structures develop and several natural orbitals are required for converged dynamics, placing the system beyond a simple two-mode description. Our results establish the interaction profile and strength, the barrier, and the initial state as distinct control parameters for engineering coherent, dephasing, recurrent, and strongly suppressed many-body tunneling dynamics of one dimensional bosons. The experimental ability to tailor soft-core interactions through Rydberg dressing provides a promising avenue for exploring interaction-engineered tunneling dynamics in ultracold atomic systems.
\end{abstract}

\section{Introduction}

Bose-Josephson junctions (BJJs) provide a paradigmatic setting for investigating superfluid dynamics and many-body effects in ultracold quantum gases~\cite{Smerzi1997, Zapata:1998, shenoy:1999, Raghavan:1999, Trombettoni:2001, Gati_2005, Albeiz:2005, Gati2007, Levy2007, Bloch2008, Sakmann2009, LeBlanc2011, Betz:2011, Spagnolli2017, Pigneur2018, Polo2018, Mennemann:2021, Ji:2022, Qiao2023, Pezze2024, Bardin2024, Vianello2025, DelPace:2025, Bernhardt:2025, Saha2026}. In the simplest realization, two weakly coupled condensates exhibit Josephson oscillations driven by tunneling between the two wells, while sufficiently strong interparticle interactions can lead to tunneling suppression when the interaction energy exceeds a critical value~\cite{Smerzi1997}. This has also been investigated in arrays of coupled Josephson junctions~\cite{Trombettoni:2001} and demonstrated experimentally~\cite{Anker:2005}. Beyond coherent Josephson oscillations, the influence of external driving, dissipation, and quantum fluctuations on the nonlinear dynamics of BJJs has been extensively explored~\cite{Levy2007,Grond2011,TrujilloMartinez2015,Lappe2018,Saha2020,Saha2023,Saha2026_res}. The dynamics of contact-interacting bosons in double-well potentials have likewise received considerable attention, both theoretically and experimentally~\cite{Smerzi1997,Milburn1997,Ananikian2006,Bausmerth:2007,MartinezGaraot2018,Saha2021}; see also the review in Ref.~\cite{Morsch2006}.

The role of spatially extended interparticle interactions in Bose-Josephson dynamics remains comparatively less explored. Previous studies have shown that the interplay between trapping geometry and the spatial structure of the interaction potential can substantially modify Josephson oscillations~\cite{AsaduzZaman:2009,Abad2011EPL,Abad2011PRA,Mazzarella2013,Saha2019}. These findings motivate a more systematic investigation of how the interaction profile, beyond its overall strength, influences the nonequilibrium many-body dynamics of a BJJ. More broadly, quantum systems with interactions that decay algebraically as $1/r^\alpha$ have attracted considerable interest, driven in part by advances in the experimental control of atomic, molecular, and optical systems~\cite{Defenu2021}. Long-range interactions have been realized or investigated in several platforms, including Rydberg atoms~\cite{Saffman}, trapped ions~\cite{Monroe:2021,Schneider:2012}, polar molecules~\cite{Carr:2009}, and dipolar quantum gases~\cite{Lahaye:2009}. Among these platforms, trapped-ion systems offer particularly versatile control over the interaction profile: suitably engineered spin couplings can exhibit tunable power-law decay, allowing the exponent $\alpha$ to be varied over a broad range~\cite{Islam,Jurcevic:2014}. Such systems have enabled experimental investigations of spin models with controlled long-range couplings~\cite{Cirac,Britton,Islam}. Rydberg-atom arrays provide a complementary platform for exploring long-range interacting quantum systems. Atoms excited to high-lying Rydberg states experience strong interparticle interactions, typically governed by van der Waals ($1/r^6$) or resonant dipole--dipole ($1/r^3$) potentials, depending on the atomic states and excitation conditions~\cite{Saffman,Browaeys2020}. The ability to arrange individual atoms in programmable geometries~\cite{Barredo2018} and control their excitation dynamics has enabled the realization of interacting spin models and the investigation of nonequilibrium many-body phenomena~\cite{Browaeys2020}. Together, these developments provide a versatile experimental setting for examining how interaction range and spatial structure influence quantum many-body dynamics.

The dynamics of long-range interacting quantum systems can differ substantially from those governed by short-range interactions, with their characteristic behavior depending sensitively on the power-law decay exponent $\alpha$~\cite{Jens:2013,Schachenmayer:2013,Vodola2014,Santos:2016,Anton:2016,Lepori2017,Defenu2018,Lerose2019,Lerose2020,Block2022,Defenu2024}. In particular, varying $\alpha$ changes the relative weight of short- and long-distance contributions to the interaction, thereby influencing collective dynamics and the development of many-body correlations. This sensitivity makes power-law interactions a promising means of controlling nonequilibrium quantum dynamics. However, a bare power-law potential, $V(r)\propto r^{-\alpha}$, is singular at short distances. To regularize this short-distance behavior while retaining a tunable power-law tail, we consider the following finite-core two-body interaction:
\begin{equation}
V_\alpha(r)=\frac{g}{r^\alpha+d^\alpha},
\label{eq:interaction}
\end{equation}
where $g$ determines the interaction strength, $d$ defines the short-distance scale, and $\alpha$ controls the spatial decay. This form provides a nonsingular interaction at $r=0$ while retaining the desired power-law behavior at large separations. 

An important point to be noticed is that varying $\alpha$ in Eq.~\eqref{eq:interaction} does not merely change the asymptotic range of the interaction. Indeed
\begin{equation}
V_\alpha(0)=\frac{g}{d^\alpha},
\end{equation}
while at large distances
\begin{equation}
V_\alpha(r)\simeq \frac{g}{r^\alpha},
\qquad r\gg d.
\end{equation}
Consequently, increasing $\alpha$ simultaneously suppresses the spatial tail and changes the strength and spatial extent of the short-distance part of the interaction. The exponent $\alpha$ should therefore be regarded as a parameter controlling the shape of the interaction kernel, rather than as a simple measure of interaction range. This clarification play a key role in understanding the many-body dynamics explored in this work.

\begin{figure}[t]
\centering
\includegraphics[width=0.8\linewidth]{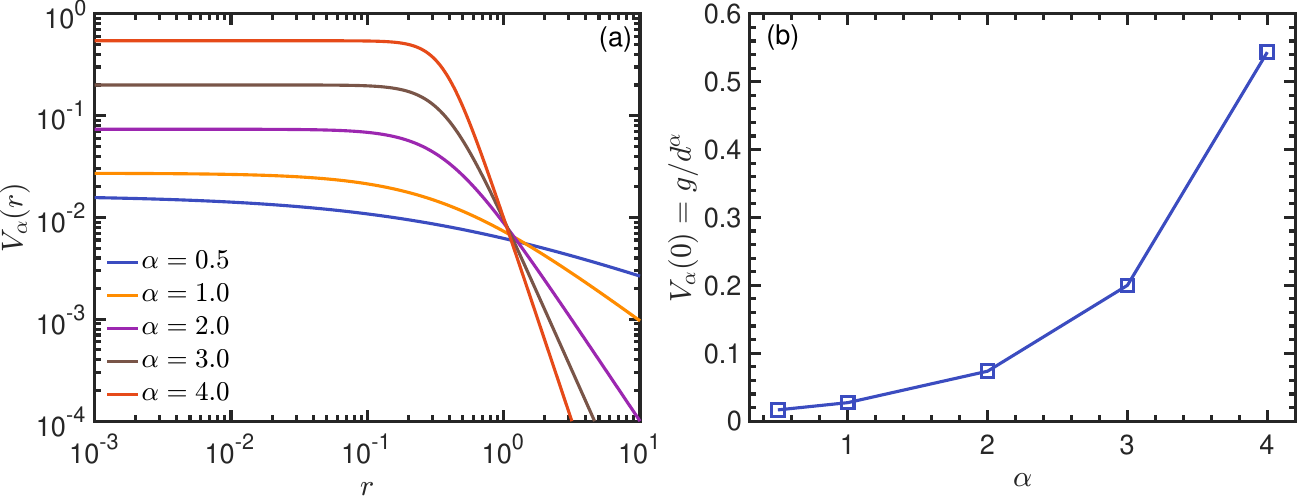}
\caption{Interaction kernel used throughout the work,
\(V_\alpha(r)=g/(r^\alpha+d^\alpha)\), with \(d^3=0.05\)~\cite{chatterjee:2018}, plotted for $g=0.01$. (a) Interaction profiles for representative values of the exponent \(\alpha\), shown on a log-log scale to highlight the asymptotic power-law behavior \(V_\alpha(r)\sim r^{-\alpha}\). (b) Corresponding short-distance behavior, illustrating the finite interaction core and its dependence on \(\alpha\).}
\label{fig1}
\end{figure}

The resulting interaction profiles are shown in Fig.~\ref{fig1}. Fig.~\ref{fig1}(a) displays $V_\alpha(r)$ for representative values of $\alpha$ on a log-log scale. The curves exhibit the expected power-law behavior at large separations, while their short-distance behavior differs substantially as $\alpha$ is varied. A magnified view of the core region in Fig.~\ref{fig1}(b) makes this variation more apparent. Although the interaction remains finite at $r=0$, its magnitude and the spatial extent over which the core influences the interaction depend on $\alpha$. Thus, even at fixed $g$, different values of $\alpha$ correspond to distinct effective interaction strengths. In the following, we therefore first keep $g$ fixed and vary $\alpha$ to examine how the spatial form of the interaction modifies the tunneling dynamics, before considering the effects of the bare interaction strength and other control parameters.

To explore the resulting many-body dynamics, we consider bosonic systems confined in a one-dimensional box \cite{Gaunt2013} and initially prepared in population-imbalanced states across a central barrier. The nonequilibrium dynamics are initiated by suddenly switching off the population-imbalance potential, thereby releasing the initially imbalanced state into the symmetric double-well potential and triggering the subsequent population transfer between the two wells. The dynamics following the quench are obtained using the multiconfigurational time-dependent Hartree method for bosons (MCTDHB) implemented in the MCTDH-X software \cite{Alon:2007,Alon:2008,Fasshauer:2016,Lode:2020,Lode:2016,lin:2020,Zhang:2024}, which allows the many-body state to evolve beyond a restricted two-mode description. We vary the interaction exponent $\alpha$, interaction strength $g$, barrier strength, and initial population imbalance to identify the distinct mechanisms governing the population dynamics.  We also compare selected MCTDHB results with the extended two-mode Bose-Hubbard (eBH) model to assess the validity of a reduced description and identify the regimes in which additional spatial orbitals become important.

Our results reveal several distinct dynamical responses controlled by different parameters. First, at weak interaction and fixed $g$, increasing $\alpha$ transforms usual Josephson oscillations (as observed with contact-like interactions) into pronounced dephasing followed by recurrent rephasing, showing the sensitivity of the dynamics to the spatial structure of the interaction kernel. Second, increasing the central barrier suppresses these recurrent population-transfer dynamics, providing direct control over the persistence of Josephson oscillations. Third, increasing the initial population imbalance drives the system toward a nearly vanishing population imbalance, resulting in relaxation of the Josephson oscillations. Finally, increasing the interaction strength progressively suppresses population-transfer fluctuations, with this suppression becoming particularly pronounced for larger $\alpha$. In the strongly interacting regime, the one-body density develops a pronounced multi-peak structure reminiscent of fermionization, while the population-imbalance dynamics become strongly suppressed. Several natural orbitals are required for a converged description, revealing a genuinely many-body regime in which strong interaction-induced fragmentation accompanies the suppression of population transfer. The comparison with the eBH model further demonstrates the limitations of a single-orbital-per-well description in this regime. As higher spatial orbitals become dynamically populated, the eBH description progressively loses its accuracy, whereas MCTDHB is shown to capture for the considered parameters the associated multiorbital structure and dynamics. Taken together, these results establish a versatile set of control knobs for engineering many-body dynamics in a BJJ. The interaction exponent, barrier strength, initial imbalance, and interaction strength provide distinct routes to selectively modify the dynamical response: from coherent Josephson oscillations to dephasing and recurrent rephasing, relaxation, and strongly suppressed population transfer.

The remainder of the paper is organized as follows. In Sec.~\ref{sec:hamiltonian}, we introduce the many-body Hamiltonian and explain the initial-state preparation and quench protocol. Sec.~\ref{sec:mctdhb} presents the multiconfigurational time-dependent Hartree method for bosons (MCTDHB) employed to solve the time-dependent many-body problem. In Sec.~\ref{sec:observables}, we define the observables used to characterize the Josephson dynamics and the underlying many-body correlations. Sec.~\ref{sec:results} presents our results, discussing the effects of the interaction exponent, interaction strength, central barrier, and initial population imbalance. Finally, Sec.~\ref{sec:conclusion} summarizes our findings and presents our conclusions. Appendix \ref{app:units} summarizes the units used in the numerics. Selected MCTDHB results from Sec.~\ref{sec:results} are compared with eBH results in Appendix \ref{appendix:BH}, and with results obtained in the case of contact interactions in Appendix \ref{app:contact}.

\section{Hamiltonian and quench protocol}\label{sec:hamiltonian}

We consider a system of $N$ interacting bosons confined in a one-dimensional trap potential $V_{\rm trap}(x)$. The many-body Hamiltonian is given by
\begin{equation}
\begin{aligned}
\hat{H} &=
\sum_{i=1}^{N}
\left[
-\frac{\hbar^{2}}{2m}\frac{\partial^{2}}{\partial x_i^{2}}
+V_{\rm trap}(x_i)
\right]
+\sum_{i<j}^{N}V_\alpha(x_i-x_j)\\
&=
\sum_{i=1}^{N}h(x_i)
+\sum_{i<j}^{N}V_\alpha(x_i-x_j),
\end{aligned}
\label{eq:Hamiltonian}
\end{equation}
where $x_i$ denotes the coordinate of the $i$-th particle and $h(x)$ is the single-particle Hamiltonian. The external trapping potential consists of three Gaussian barriers located at $x=0$ and $x=\pm L$,
\begin{equation}
\begin{aligned}
V_{\rm ext}(x)=
V_1\exp\left(-\frac{x^2}{\sigma^2}\right)
+V_0\exp\left[-\frac{(x-L)^2}{\sigma^2}\right]
+V_0\exp\left[-\frac{(x+L)^2}{\sigma^2}\right].
\end{aligned}
\label{eq:external_potential}
\end{equation}
The two outer Gaussian barriers provide an effective implementation of boundary conditions, while the central barrier of strength $V_1$ separates the system into two weakly coupled reservoirs. For sufficiently large $V_0$, the resulting potential provides box-like confinement, as illustrated in Fig.~\ref{fig:trap_pot}. Throughout this work, we use dimensionless units with $\hbar=m=1$ (see Appendix \ref{app:units}), fix the Gaussian width to $\sigma=0.1$, and consider a box of total length $2L=8$, corresponding to $x\in[-4,4]$. 

\begin{figure}[b]
\centering
\includegraphics[width=0.8\linewidth]{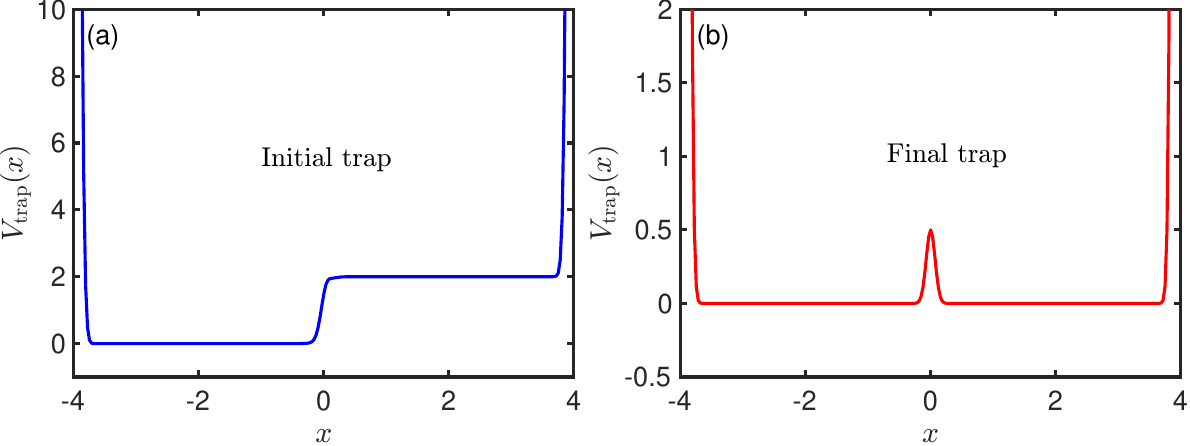}
\caption{Schematic representation of the quench protocol for $V_1=0.5$ and $V_{\rm imb}=2.0$. (a) Initial trapping potential, $V_{\rm trap}(x)=V_{\rm ext}(x)+V_{\rm offset}(x)$, where the auxiliary offset potential creates a population imbalance between the two reservoirs. (b) Symmetric trapping potential, $V_{\rm ext}(x)$, following the sudden removal of the offset at $t=0$. The central Gaussian barrier remains unchanged during the quench and controls the coupling between the two reservoirs.}
\label{fig:trap_pot}
\end{figure}

To prepare an initial population imbalance between the two reservoirs, we introduce an auxiliary offset potential,
\begin{equation}
V_{\rm offset}(x)=\frac{V_{\rm imb}}{2}
\left[
\tanh\left(\frac{x}{\sigma}\right)
+\tanh\left(\frac{L-x}{\sigma}\right)
\right],
\label{eq:offset_potential}
\end{equation}
where $V_{\rm imb}$ controls the magnitude of the initial imbalance. The system is initially prepared in the ground state of the combined potential
\begin{equation}
V_{\rm trap}(x)=V_{\rm ext}(x)+V_{\rm offset}(x),
\label{eq:initial_trap}
\end{equation}
obtained by imaginary-time propagation. At $t=0$, the offset amplitude is switched off and the system evolves in real time under $V_{\rm ext}(x)$ alone. The initial and post-quench trapping potentials are illustrated in Figs.~\ref{fig:trap_pot}(a) and \ref{fig:trap_pot}(b), respectively. Thus, the quench removes the initial potential asymmetry while leaving the central barrier and interaction parameters unchanged.

The subsequent nonequilibrium dynamics are governed by the time-dependent many-body Schrödinger equation,
\begin{equation}
i\hbar\frac{\partial}{\partial t}|\Psi(t)\rangle
=\hat{H}|\Psi(t)\rangle.
\label{eq:TDSE}
\end{equation}
Unless stated otherwise, the parameter ranges explored in this work are $\alpha\in[0.5,4.0]$, $g\in[0.01,0.5]$, $V_1\in[0.5,3.0]$, and $V_{\rm imb}\in[0.2,2.0]$.

\section{Multiconfigurational time-dependent Hartree method for bosons (MCTDHB)}\label{sec:mctdhb}

To investigate the nonequilibrium many-body dynamics of the Hamiltonian \eqref{eq:Hamiltonian}, we employ the MCTDHB. Based on the time-dependent variational principle, it provides a systematically improvable description of interacting bosons beyond the single-orbital mean-field approximation. The method expands the many-body wavefunction in a time-dependent basis of permanents constructed from a set of $M$ self-consistently optimized single-particle orbitals. Both the expansion coefficients and the orbitals evolve in time, allowing the basis to adapt to the correlations generated during the dynamics. The many-body wavefunction is expressed as
\begin{equation}
|\Psi(t)\rangle =
\sum_{\vec{n}} C_{\vec{n}}(t)|\vec{n};t\rangle,
\label{many_body_wf}
\end{equation}
where $\vec{n}=(n_1,n_2,\ldots,n_M)$ denotes an occupation-number configuration satisfying $\sum_{k=1}^{M}n_k=N$. The permanents forming the basis are given by
\begin{equation}
|\vec{n};t\rangle =
\prod_{k=1}^{M}
\frac{[\hat b_k^\dagger(t)]^{n_k}}{\sqrt{n_k!}}|0\rangle,
\label{many_body_wf_2}
\end{equation}
where $|0\rangle$ is the vacuum state and $\hat b_k^\dagger(t)$ creates a boson in the time-dependent orbital $\psi_k(x,t)$. The coefficients $C_{\vec n}(t)$ and orbitals $\psi_k(x,t)$ are determined variationally by applying the Dirac-Frenkel variational principle to the many-body action. This coupled evolution enables the method to follow the redistribution of particles among orbitals and the development of many-body correlations throughout the quench dynamics. 

An important feature of MCTDHB is its ability to describe fragmentation, in which more than one natural orbital acquires a significant occupation. This is particularly relevant to the present work, where the finite-core power-law interaction can generate substantial changes in the spatial structure and correlations of the evolving state. Unlike a fixed-orbital description, the time-adaptive basis can respond to these changes, making MCTDHB suitable for exploring the crossover from nearly coherent Josephson oscillations to dephasing, recurrent dynamics, and strongly correlated regimes.

The method becomes exact in the limit $M\rightarrow\infty$. In practical calculations, a finite orbital number is sufficient provided that the relevant observables are converged with respect to $M$. We therefore assess convergence by increasing the number of time-dependent orbitals and monitoring the stability of the population imbalance, natural-orbital occupations, and many-body correlation measures. The variational framework and equations of motion underlying MCTDHB are described in Refs.~\cite{TDVM81,variational1,variational3,variational4}; applications to ground-state and nonequilibrium properties of ultracold bosonic systems with long-range interaction can be found in Refs.~\cite{Chakrabarti_2025,Molignini2024,Molignini-2025,Molignini:2025,Fischer2015,Bera2019,Roy2024}.

\section{Observables}\label{sec:observables}

To characterize the nonequilibrium many-body dynamics following the quench, we monitor the population imbalance, the frequency of the oscillations, and the occupation of the natural orbitals. These observables provide complementary information about particle transfer between the two reservoirs, the characteristic timescales of the dynamics, and the development of many-body correlations.

\textbf{Population imbalance.} The population imbalance between the left and right wells is
\begin{equation}
z(t)=\frac{N_L(t)-N_R(t)}{N},
\end{equation}
where $N_L(t)$ and $N_R(t)$ denote the instantaneous particle numbers in the left and right wells, respectively, and $N=N_L(t)+N_R(t)$ is the total particle number. The time evolution of $z(t)$ characterizes the population transfer between the two reservoirs and provides a direct measure of the dynamics.

\textbf{Oscillation frequency.} The characteristic frequency of the population-transfer dynamics is extracted from the Fourier spectrum of $z(t)$. Denoting the Fourier transform of the population imbalance by $z(\omega)$, we define the power spectrum as $|z(\omega)|^2$. For ease of comparison between different interaction parameters, we normalize the spectrum by dividing it by its maximum value. The dominant frequency $\omega_\alpha$ is identified from the position of the principal peak in the normalized spectrum for a given interaction exponent $\alpha$. The associated characteristic timescale is then defined as $\tau_\alpha=2\pi/\omega_\alpha$. This quantity provides a measure of the dominant oscillatory timescale, including regimes in which the population imbalance exhibits dephasing and recurrent dynamics.

\textbf{One-body reduced density matrix and natural-orbital occupations.} The one-body reduced density matrix is defined as
\begin{equation}
\rho^{(1)}(x,x';t)=\langle\Psi(t)|\hat{\Psi}^{\dagger}(x')\hat{\Psi}(x)|\Psi(t)\rangle,
\end{equation}
where $\hat{\Psi}^{\dagger}(x)$ and $\hat{\Psi}(x)$ are the bosonic field creation and annihilation operators, respectively. Diagonalization of this matrix yields its spectral decomposition,
\begin{equation}
\rho^{(1)}(x,x';t)=\sum_{j=1}^{M}n_j(t)\phi_j(x,t)\phi_j^*(x',t),
\end{equation}
where $\phi_j(x,t)$ are the natural orbitals and $n_j(t)$ are their corresponding occupations. The natural orbitals are ordered according to decreasing occupation, such that $n_1(t)\geq n_2(t)\geq\cdots\geq n_M(t)$, with $\sum_{j=1}^{M}n_j(t)=N$.

The distribution of particles among the natural orbitals provides a measure of the degree of condensation and fragmentation during the nonequilibrium evolution. A state dominated by a single natural orbital, with $n_1\simeq N$, is approximately fully condensed, whereas significant occupations of two or more natural orbitals indicate fragmentation \cite{Penrose1956, Leggett2001}. Monitoring the time dependence of the natural-orbital occupations therefore allows us to assess the development of multiorbital correlations induced by the interaction and the quench.

\section{Results and Discussion}\label{sec:results}

We investigate how the interaction profile, interaction strength, central barrier, and initial population imbalance govern the nonequilibrium many-body dynamics of the considered BJJ. The results are organized around five distinct dynamical responses: (i) interaction-structure-induced dephasing and recurrent rephasing, (ii) suppression of recurrent dynamics by increasing the central barrier, (iii) relaxation toward a nearly vanishing population imbalance as the initial imbalance is increased, (iv) suppression of population-transfer fluctuations with increasing interaction strength, and (v) the emergence of multiorbital correlations in the strongly interacting regime. The first three investigations, concerning the interaction profile, barrier strength, and initial imbalance, are performed for $N=10$ bosons. The interaction-strength dependence is examined separately, while the strongly interacting regime and its associated multiorbital correlations are studied for $N=8$ bosons because of the increased computational demands.

\begin{figure}[t]
\centering
\includegraphics[width=11cm]{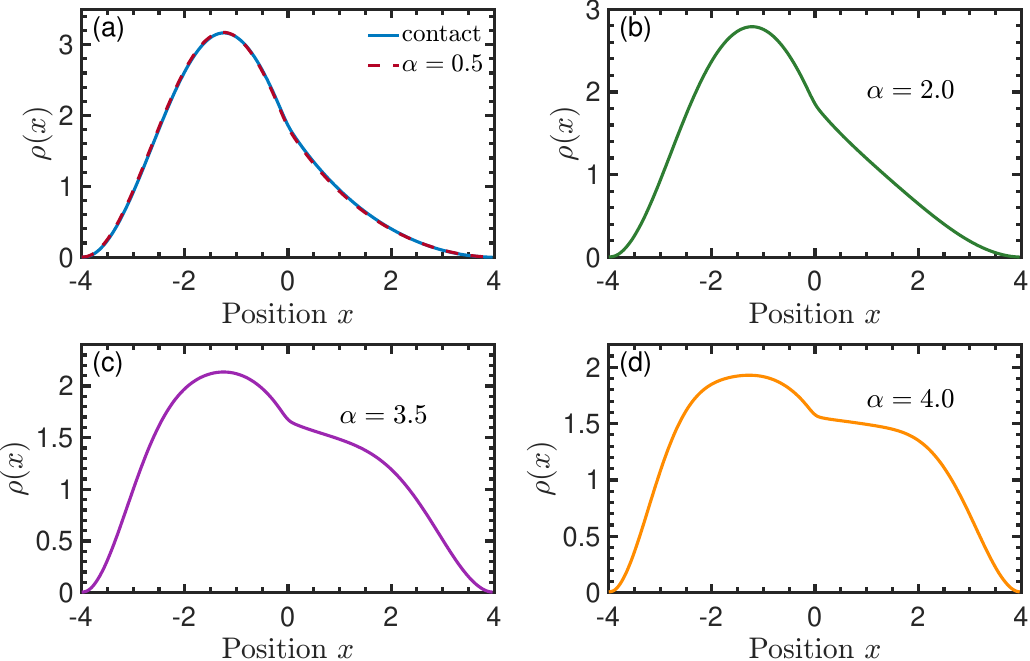}
\caption{One-body density distributions for different values of the interaction exponent $\alpha$, with the interaction strength, the central barrier, and initial offset amplitude fixed at $g=0.01$, $V_1=0.5$, and $V_{\rm imb}=0.2$, respectively. The contact-interaction and $\alpha=0.5$ profiles are nearly indistinguishable. Increasing $\alpha$ progressively modifies the density profile, from an asymmetric, predominantly left-localized distribution to a broader, structured profile extending across both wells, with a kink near the central barrier at $x=0$.}
\label{fig:density}
\end{figure}

\subsection{Interaction structure-controlled tunneling dynamics}

We first examine how the interaction profile influences the spatial distribution of the bosons and the ensuing tunneling dynamics. The results presented in this subsection are obtained using $M=4$ time-dependent orbitals within MCTDHB. Fig.~\ref{fig:density} shows the one-body density for different values of the interaction exponent $\alpha$ at fixed $g=0.01$, keeping the central barrier and offset amplitude fixed at $V_1=0.5$ and $V_{\rm imb}=0.2$, respectively. The contact-interaction and $\alpha=0.5$ profiles are nearly indistinguishable, indicating that the latter provides a contact-like reference in the parameter regime considered. For such weak interaction, the density is predominantly localized in the left well and exhibits an asymmetric, approximately Gaussian profile. As $\alpha$ increases, the density develops a progressively more pronounced kink around the central barrier at $x=0$, and the initially localized profile gives way to a broader, structured distribution extending across the trap, with a spatially-fragmented-Gaussian-like appearance. In fact, larger $\alpha$ leads to a stronger effective repulsion, forcing the bosons to spread across the whole trap despite the offset potential.

\begin{figure}[t]
\centering
\includegraphics[width=11cm]{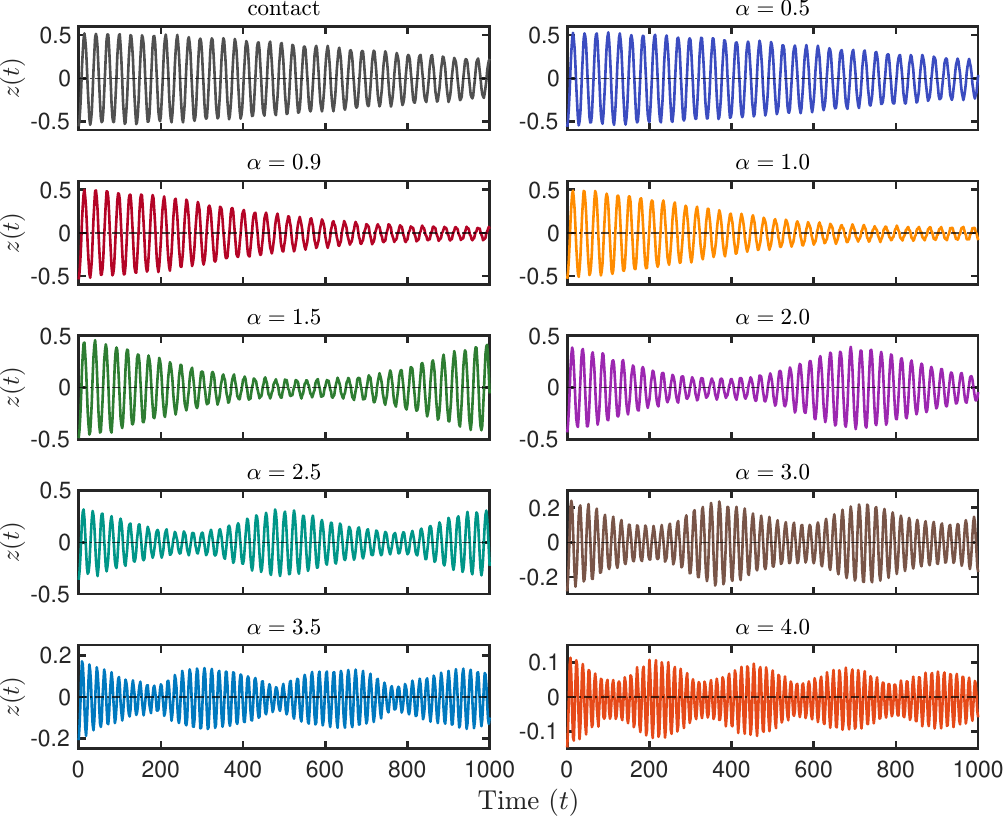}
\caption{Interaction-dependent population-imbalance dynamics. The population imbalance $z(t)$ is shown for $g=0.01$, $V_1=0.5$, $V_{\rm imb}=0.2$, and $N=10$ for the contact interaction and different values of the interaction exponent $\alpha$. Increasing $\alpha$ progressively modifies the phase evolution of the oscillations, leading to dephasing followed by recurrent rephasing.}
\label{fig2}
\end{figure}

Next, we investigate how the interaction structure controls the nonequilibrium dynamics following the sudden removal of the initial offset potential. Fig.~\ref{fig2} presents the time evolution of the population imbalance $z(t)$ for the contact interaction and a sequence of finite-core power-law interactions, illustrating the progressive changes in the dynamical response as the interaction exponent $\alpha$ is varied. For the contact interaction, the system exhibits regular Josephson oscillations. The dynamics for $\alpha=0.5$ closely follow this behavior, providing a natural contact-like reference within the finite-core power-law family. As $\alpha$ increases, the oscillations progressively lose their regularity, with successive maxima and minima shifting relative to the contact-like oscillatory pattern. This dephasing becomes increasingly pronounced at larger $\alpha$. After this initial transient, coherent oscillatory structures re-emerge at later times, indicating recurrent rephasing. This systematic evolution demonstrates that the interaction exponent provides a control parameter for modifying the Josephson dynamics through changes in the spatial structure of the interaction kernel. Since the interaction strength and trapping potential are unchanged throughout this comparison, the observed changes originate from the modification of the interaction profile rather than from a variation in the overall interaction strength. Thus, tuning $\alpha$ offers a route to engineering the dynamical response of the junction, from regular Josephson oscillations to pronounced dephasing and recurrent rephasing.

\begin{figure}[t]
\centering
\includegraphics[width=11cm]{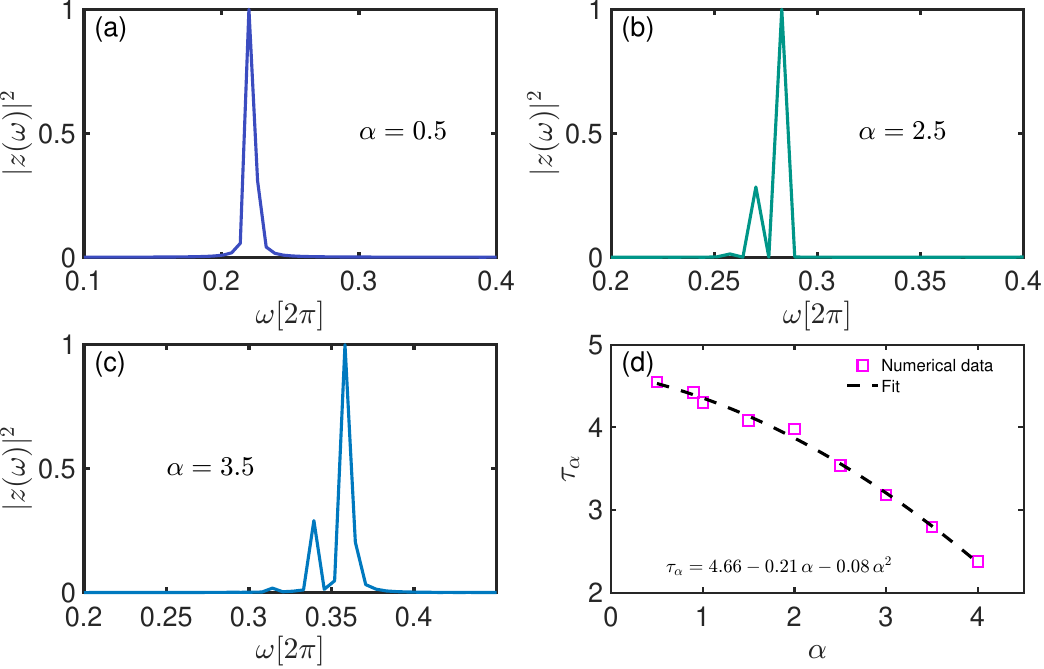}
\caption{Frequency-domain analysis of the interaction-dependent Josephson dynamics. (a)-(c) Fourier spectra of $z(t)$ for representative interaction exponents $\alpha=0.5$, $2.5$, and $3.5$. The dominant spectral peak shifts toward larger $\omega$ with increasing $\alpha$, while additional closely spaced features remain strongly overlapped with the main peak. (d) Characteristic timescale $\tau_\alpha=2\pi/\omega_\alpha$ extracted from the position $\omega_\alpha$ of the dominant peak as a function of $\alpha$. The dashed line indicates the quadratic fitting expression $\tau_\alpha \simeq 4.66-0.21\alpha-0.08\alpha^2$.}
\label{fig3}
\end{figure}

To further characterize the interaction-exponent effects on the dynamics, we examine the frequency content of the population imbalance $z(t)$. The Fourier spectra provide a complementary perspective on the time-domain evolution and allow us to identify the dominant frequency scale of the Josephson oscillations. For each interaction exponent $\alpha$, we define the dominant frequency $\omega_\alpha$ as the position of the principal peak in the Fourier spectrum of $z(t)$ and the corresponding characteristic timescale as $\tau_\alpha=2\pi/\omega_\alpha$. Figs.~\ref{fig3}(a)-(c) show the Fourier spectra for representative values of the interaction exponent, $\alpha=0.5$, $2.5$, and $3.5$. In all three cases, the spectrum is characterized by a pronounced dominant peak. As $\alpha$ increases, this peak shifts towards higher frequencies, from approximately $\omega\simeq0.22$ at $\alpha=0.5$ to $\omega\simeq0.38$ at larger exponents. For larger $\alpha$, a secondary peak also becomes discernible close to the dominant one. We use the principal peak to define the dominant frequency scale, while the additional peak points to a secondary contribution to the population-imbalance dynamics. The shift of the dominant peak towards higher frequencies implies a reduction in the characteristic timescale $\tau_\alpha$. Thus, increasing the interaction exponent modifies not only the temporal pattern of dephasing and recurrent rephasing observed in $z(t)$, but also the characteristic timescale of the population-transfer dynamics. Notably, the dephasing observed in the time domain does not correspond to a broad distribution of well-separated spectral peaks. Instead, the spectra retain a pronounced dominant frequency, accompanied by a weaker secondary contribution at larger $\alpha$. The characteristic timescale $\tau_\alpha$ as a function of $\alpha$ is shown in Fig.~\ref{fig3}(d). It decreases systematically with increasing $\alpha$, indicating an acceleration of the dominant oscillatory dynamics as the interaction profile is modified. The dependence is well described by a quadratic fit, $\tau_\alpha \simeq 4.66 - 0.21\alpha - 0.08\alpha^2$, which quantifies the nonlinear variation of the characteristic timescale with the interaction exponent. Taken together, the time-domain and frequency-domain analyses demonstrate that the interaction exponent provides control over both the temporal organization and the characteristic frequency scale of the Josephson dynamics.

\subsection{Barrier-controlled suppression of recurrent dynamics}

\begin{figure}[t]
\centering
\includegraphics[width=11cm]{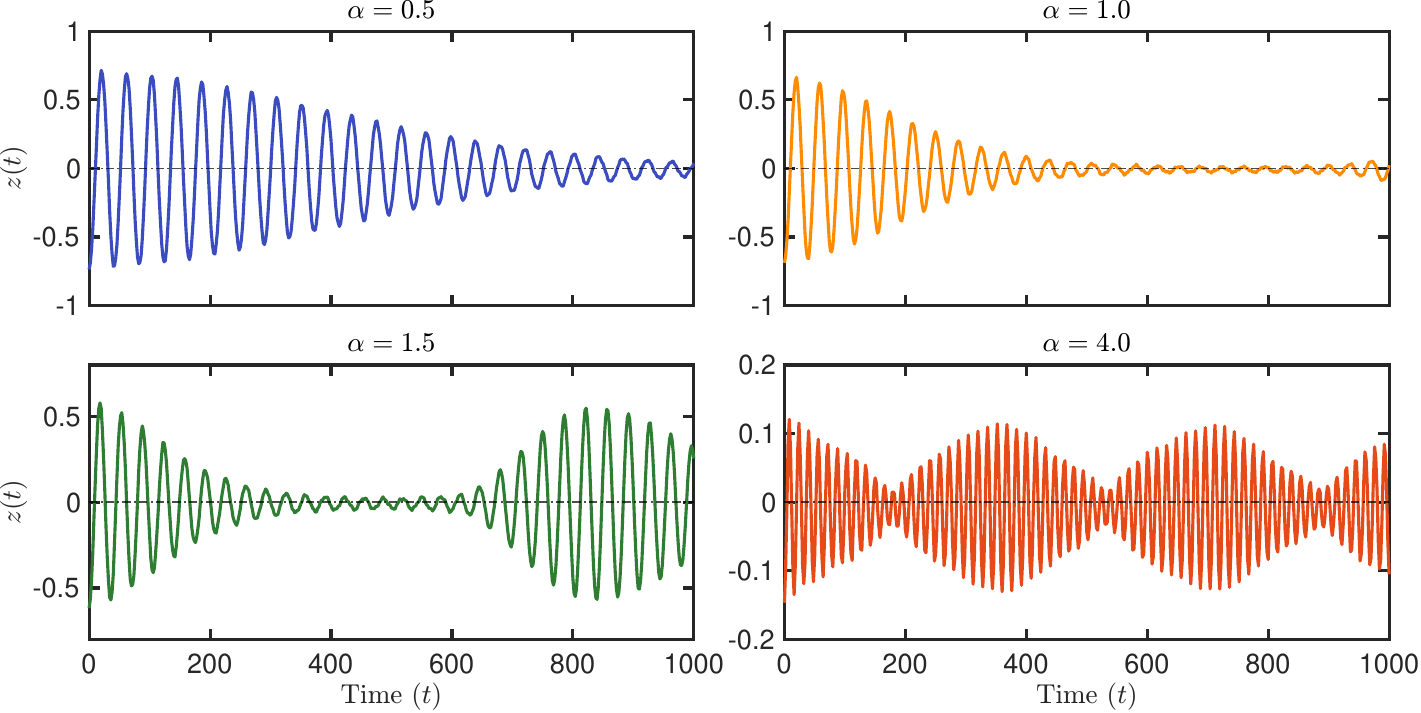}
\caption{Barrier-controlled population-imbalance dynamics for $g=0.01$, $V_1=3.0$, $V_{\rm imb}=0.2$, and $N=10$, for the representative interaction exponents $\alpha=0.5$, $1.0$, $1.5$, and $4.0$. The dynamics shows strong damping at $\alpha=1.0$, with the imbalance approaching a nearly stationary value close to zero. At larger $\alpha$, oscillatory and recurrent features re-emerge on the time scale considered, and the number of visible oscillation cycles increases towards $\alpha=4.0$.}
\label{fig4}
\end{figure}

We next investigate the effects of the central barrier. The results presented in this subsection are obtained using $M=4$ time-dependent orbitals within MCTDHB. Fig.~\ref{fig4} shows the population-imbalance evolution for the stronger barrier, $V_1=3.0$, while keeping the interaction strength and initial offset amplitude fixed. This result can be compared with the corresponding dynamics for the weaker barrier, $V_1=0.5$, presented in Fig.~\ref{fig2}, to assess how increasing the central barrier modifies the interaction-exponent-dependent Josephson dynamics.

For the stronger barrier, $V_1=3.0$, four representative interaction exponents are considered: $\alpha=0.5$, $1.0$, $1.5$, and $4.0$. For $\alpha=0.5$, the population-transfer oscillations are damped, and for $\alpha=1.0$ the damping becomes more pronounced and the imbalance approaches a nearly stationary value close to zero. At $\alpha=1.5$, oscillatory structures begin to re-emerge. For $\alpha=4.0$, a larger number of oscillation cycles becomes visible, together with signatures of dephasing and recurrent rephasing. Nevertheless, the number of observable cycles remains smaller than that obtained for the weaker barrier.

These results show that increasing the central barrier does not completely suppress the interaction-induced dephasing and recurrent rephasing. Rather, it modifies their temporal development, reducing the persistence of population transfer and limiting the number of observable oscillation cycles. The central barrier therefore acts as an additional control parameter for tuning the Josephson response, complementing the control exerted by the interaction exponent over the temporal organization of the dynamics.

\subsection{Initial-imbalance-controlled relaxation dynamics}

\begin{figure}[b]
\centering
\includegraphics[width=11cm]{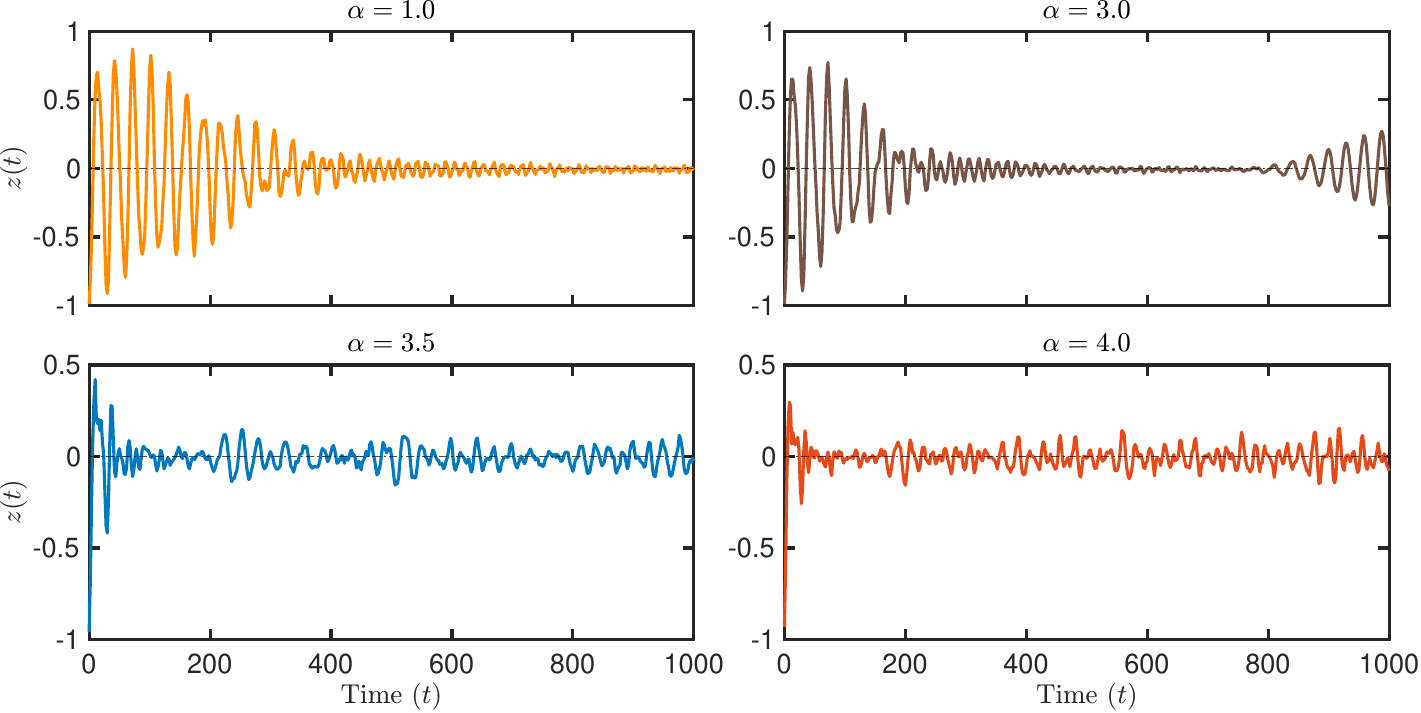}
\caption{Initial-imbalance-controlled relaxation dynamics for $g=0.01$, $V_1=0.5$, $V_{\rm imb}=2.0$, and $N=10$, for the representative interaction exponents $\alpha=1.0$, $3.0$, $3.5$, and $4.0$. Increasing $\alpha$ progressively suppresses the initial transient and population-imbalance oscillations. For $\alpha=3.5$ and $4.0$, only a small initial transient is observed, followed by fast relaxation towards $z(t)\simeq 0$.}
\label{fig5}
\end{figure}

We next examine how the initial population imbalance influences the subsequent nonequilibrium dynamics. The results presented in this subsection are obtained using $M=4$ time-dependent orbitals within MCTDHB. Keeping the interaction strength and central barrier fixed at $g=0.01$ and $V_1=0.5$, respectively, we examine the dynamics for the larger initial offset amplitude, $V_{\rm imb}=2.0$, as shown in Fig.~\ref{fig5}. The resulting population-imbalance evolution can be compared with the weakly imbalanced case, $V_{\rm imb}=0.2$, shown in Fig.~\ref{fig2}.

The larger initial imbalance leads to a qualitatively different dynamical response. For $\alpha=1.0$, the evolution exhibits an initial transient followed by strong damping of the population-imbalance oscillations. As $\alpha$ increases to $3.0$, the initial transient becomes less evident, and the population imbalance approaches $z(t)\simeq 0$ more rapidly. For $\alpha=3.5$ and $4.0$, only a very small initial transient remains, after which the imbalance rapidly relaxes towards zero and stays close to the balanced population state, with only small residual fluctuations.

These results show that the initial imbalance provides an additional tool of controlling the Josephson dynamics. In particular, for a strongly imbalanced initial state, increasing the interaction exponent progressively suppresses the initial transient and subsequent population transfer, driving the system towards a nearly balanced state. This response contrasts with the dephasing and recurrent rephasing observed for the weakly imbalanced preparation, highlighting the interplay between the interaction profile and the initial state in shaping the nonequilibrium dynamics.

To quantify the relaxation towards the balanced state, we introduce the root-mean-square (RMS) population imbalance,
\begin{equation}
A_0=\left[\frac{1}{T}\int_0^T z^2(t)\,dt\right]^{1/2},
\label{eq:A0}
\end{equation}
where $T$ is the total evolution time considered. Unlike an RMS fluctuation measured relative to the time-averaged imbalance, $A_0$ quantifies the overall magnitude of the population imbalance relative to $z=0$. A reduction in $A_0$ therefore indicates that the dynamics become increasingly concentrated around the balanced state.

\begin{figure}[t]
\centering
\includegraphics[width=7cm]{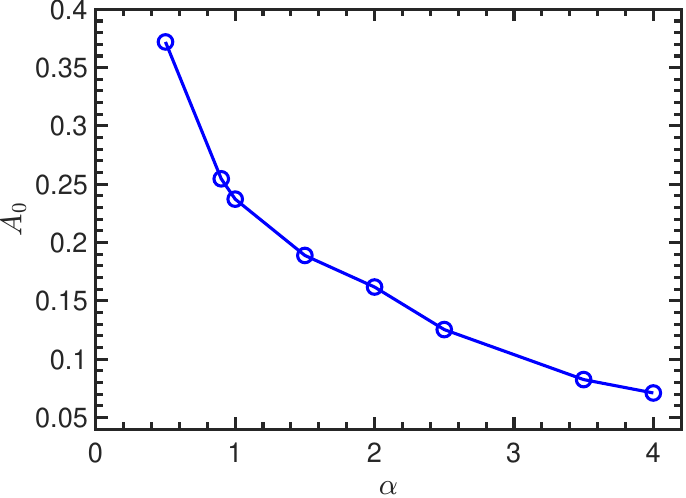}
\caption{Quantification of relaxation for a strongly imbalanced initial state. The root-mean-square population imbalance $A_0$ is shown as a function of the interaction exponent $\alpha$ for $g=0.01$, $V_1=0.5$, $V_{\rm imb}=2.0$, and $N=10$. The decrease in $A_0$ with increasing $\alpha$, reaching approximately $0.07$ at $\alpha=4.0$, quantifies the increasing concentration of the population imbalance around the balanced state $z=0$.}
\label{fig6}
\end{figure}

For the strongly imbalanced initial condition, $A_0$ decreases non-linearly with increasing $\alpha$, as shown in Fig.~\ref{fig6}. The strongest suppression occurs at $\alpha=4.0$, where $A_0$ reaches approximately $0.07$. This is consistent with the time-domain analysis: the initial transient becomes progressively weaker, and the population imbalance remains close to zero at later times. For $\alpha=3.0$, $3.5$, and $4.0$, the long-time dynamics are qualitatively similar, with the main differences arising in the magnitude and duration of the initial transient.

Taken together, these results demonstrate that the influence of the interaction exponent depends strongly on the initial population imbalance. For weak initial imbalance, varying $\alpha$ primarily modifies the temporal organization of population transfer, producing dephasing and recurrent rephasing. For strong initial imbalance, increasing $\alpha$ instead suppresses the transient dynamics and promotes relaxation towards the balanced state.

\subsection{Interaction-induced suppression of population transfer}

\begin{figure}[t]
\centering
\includegraphics[width=12cm]{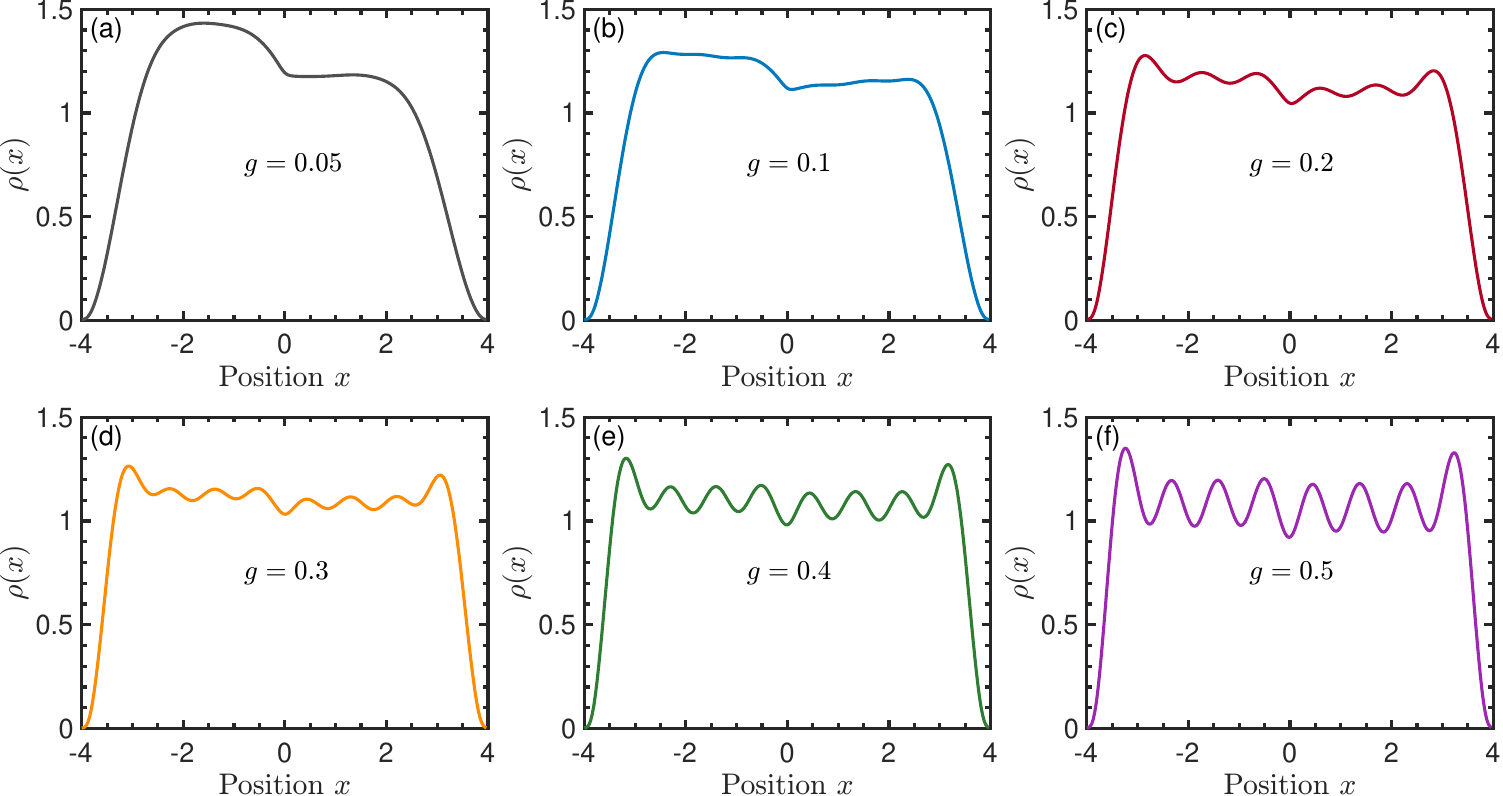}
\caption{Interaction-strength dependence of the one-body density. The one-body density $\rho(x)$ is shown for $\alpha=3.0$, $V_1=0.5$, $V_{\rm imb}=0.2$, and $N=8$, for increasing interaction strengths $g$. With increasing $g$, the density develops a multi-peak structure, with approximately four peaks located in each well at the strongest interaction considered.}
\label{fig7}
\end{figure}

We next investigate the role of the interaction strength $g$ in the many-body dynamics. To isolate its effect, we fix the interaction exponent at $\alpha=3.0$ and consider $V_1=0.5$ and $V_{\rm imb}=0.2$. The interaction strength is varied from $g=0.01$ to $g=0.5$. The MCTDHB calculations employ $M=4$ orbitals for very weak interactions, increasing to $M=6$ at intermediate interaction strengths and to $M=10$ for the strongest interaction considered, $g=0.5$. This increase in the orbital number allows us to describe the increasingly structured density profiles and associated many-body correlations as the interaction strength grows. The corresponding ground-state density profiles for different interaction strengths are shown in Fig.~\ref{fig7}. Owing to the increased computational demands of these multiorbital calculations, here we consider $N=8$ bosons.

For the weakly interacting case, $g=0.01$, the ground-state density exhibits a predominantly Gaussian-like profile with a kink at $x=0$, reflecting the presence of the central barrier. As $g$ increases, the ground-state density undergoes a clear structural transformation. Additional spatial peaks progressively develop, and the initially weak secondary structure becomes increasingly pronounced. At sufficiently strong interaction, the density exhibits a clear multi-peak structure distributed across both wells. In particular, for $g=0.5$, eight distinct density maxima can be resolved, with approximately four peaks in each well. The appearance of a number of distinct density maxima equal to the number of particles indicates a substantial restructuring of the ground state, consistent with the increasing importance of interparticle correlations in the strongly interacting regime. 

\begin{figure}[b]
\centering
\includegraphics[width=11cm]{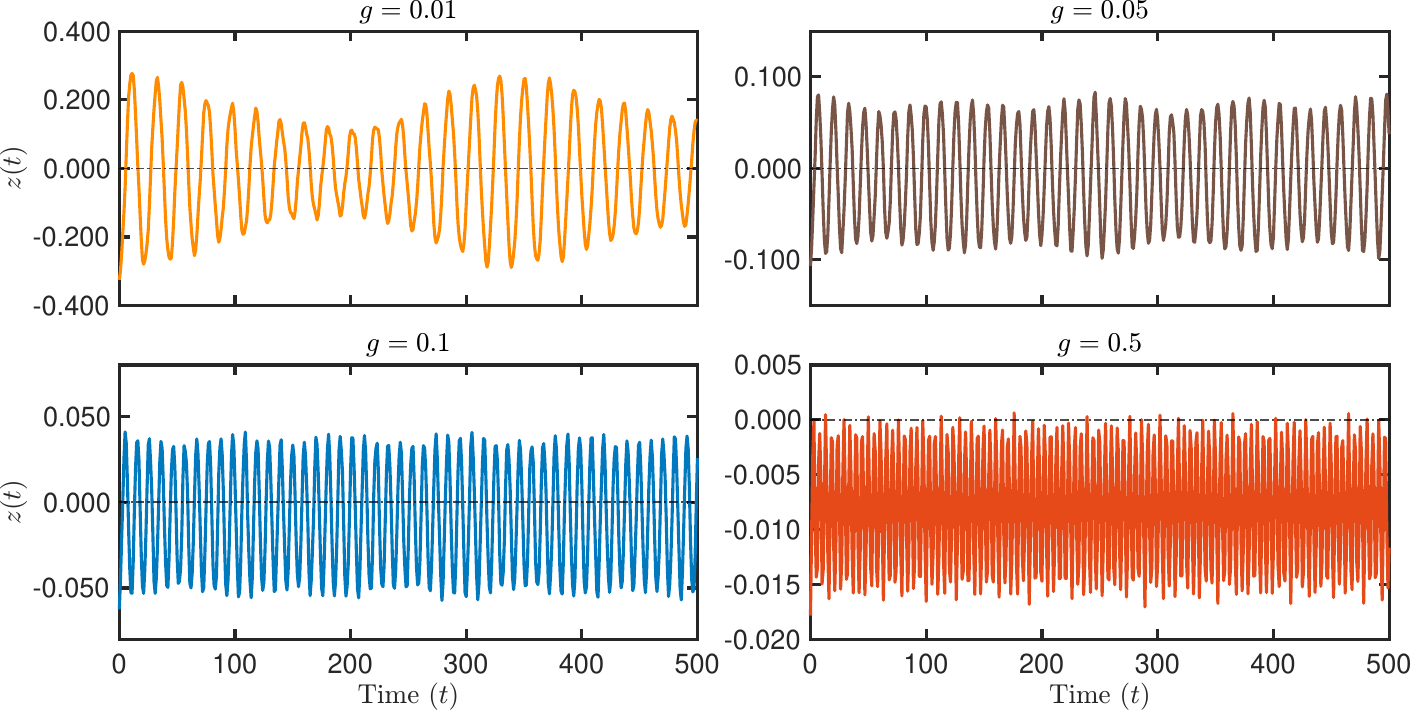}
\caption{Interaction-strength dependence of the population-imbalance dynamics. The population imbalance $z(t)$ is shown for $\alpha=3.0$, $V_1=0.5$, $V_{\rm imb}=0.2$, and $N=8$, for interaction strengths $g=0.01$, $0.05$, $0.1$, and $0.5$. Increasing $g$ progressively suppresses the amplitude of the oscillations, indicating a reduction in population transfer between the two wells.}
\label{fig8}
\end{figure}

We next examine how this interaction-induced restructuring of the initial state is reflected in the subsequent quench dynamics and, in particular, in the suppression of population-transfer fluctuations. The corresponding population-imbalance dynamics are shown in Fig.~\ref{fig8}. As the interaction strength increases, the amplitude of the oscillations is progressively suppressed, indicating a substantial reduction in population transfer between the two wells. This suppression becomes extreme in the strongly interacting regime, where $z(t)$ exhibits only small residual fluctuations around zero. Thus, the strongly interacting regime is characterized by the coexistence of a highly structured spatial density and strongly suppressed population transfer, highlighting the distinct roles of interaction strength in shaping both the spatial organization and the dynamical response of the junction.

\begin{figure}[t]
\centering
\includegraphics[width=7cm]{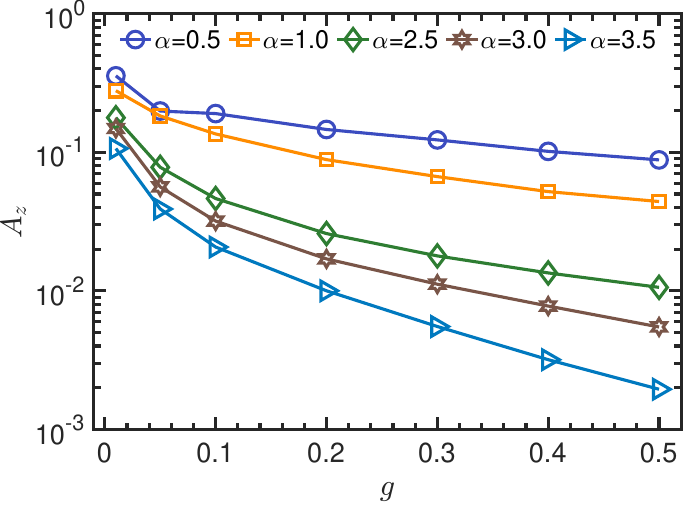}
\caption{Interaction-strength dependence of the population-imbalance fluctuations. The root-mean-square fluctuation amplitude $A_z$ is shown as a function of the interaction strength $g$ for different values of the interaction exponent $\alpha$, with $V_1=0.5$, $V_{\rm imb}=0.2$, and $N=8$. The overall decrease of $A_z$ with increasing $g$ demonstrates the suppression of population-transfer fluctuations, with this suppression becoming more pronounced at larger $\alpha$.}
\label{fig9}
\end{figure}

To quantify the suppression of population-transfer fluctuations, we introduce the RMS fluctuation of the population imbalance about its time-averaged value,
\begin{equation}
A_z=
\left[
\frac{1}{T}\int_0^T
\left(z(t)-\bar z\right)^2\,dt
\right]^{1/2},
\label{eq:Az}
\end{equation}
where $\bar z=T^{-1}\int_{0}^T z(t)\, dt$ is the time-averaged population imbalance. The resulting dependence of $A_{z}$ on the interaction strength $g$ is shown in Fig.~\ref{fig9}. This quantity measures the amplitude of temporal fluctuations around the mean imbalance and therefore provides a measure of the extent of population-transfer fluctuations between the two wells.

At weak interaction, the population imbalance exhibits substantial temporal variations, resulting in relatively large values of $A_z$. As the interaction strength increases, these variations become progressively suppressed. For the contact interaction and $\alpha=0.5$, $A_z$ decreases from values of order unity at weak interaction to approximately $10^{-1}$ at stronger interaction. For larger $\alpha$, the suppression becomes substantially stronger, with $A_z$ reaching values of order $10^{-3}$ for $\alpha=3.5$ at $g=0.5$. This indicates that in the strongly interacting regime the population imbalance explores a much narrower range around its time-averaged value. Although $A_z$ does not decrease strictly monotonically with $g$ for every interaction exponent, the overall trend demonstrates a strong suppression of population-imbalance fluctuations with increasing interaction strength. This effect is particularly pronounced for larger $\alpha$, showing that the interaction strength and the spatial structure of the interaction kernel act together to constrain the population-transfer dynamics.

\begin{figure}[b!]
\centering
\includegraphics[width=11cm]{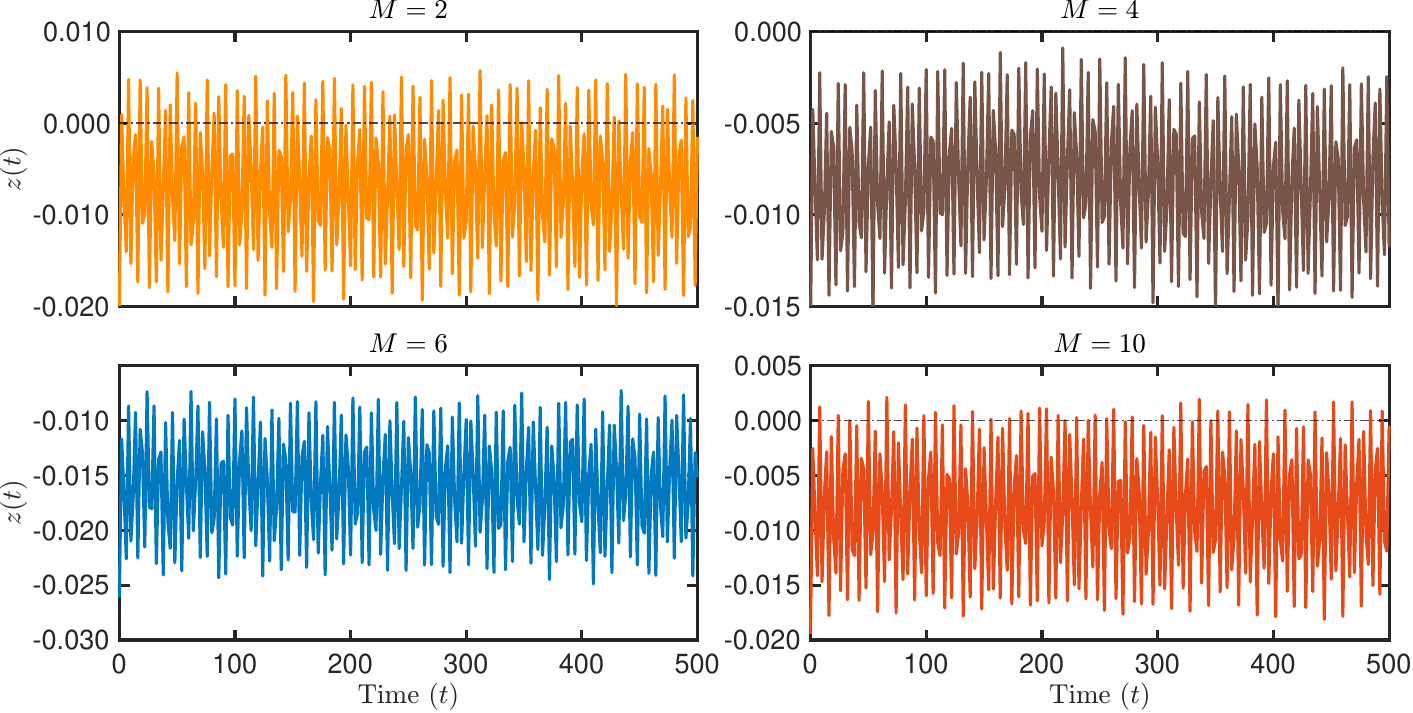}
\caption{Orbital-number dependence of the population-imbalance dynamics for $\alpha=3.0$, $g=0.5$, $V_1=0.5$, $V_{\rm imb}=0.2$, and $N=8$. The population imbalance $z(t)$ is shown for $M=2$, $4$, $6$, and $10$ time-dependent orbitals. Despite the increasing multiorbital character of the many-body description, the overall temporal structure of $z(t)$ remains nearly unchanged, indicating the robustness of this observable against the inclusion of additional orbitals in the parameter regime considered.}
\label{fig10}
\end{figure}

\subsection{Many-body fragmentation and population dynamics}

We finally examine the role of higher orbitals in the strongly interacting regime. We fix $\alpha=3.0$, $g=0.5$, $V_1=0.5$, and $V_{\rm imb}=0.2$, and perform calculations with $M=2$, $4$, $6$, and $10$ time-dependent orbitals. Here, $M$ denotes the number of variational single-particle orbitals used to construct the many-body wavefunction. Increasing $M$ enlarges the available many-body configuration space, allowing the system to develop a richer multiorbital structure and capture correlations that may be inaccessible with a smaller orbital basis. Comparing different values of $M$ therefore allows us to assess the sensitivity of the observables to the size of the variational basis. This parameter regime corresponds to the strongest interaction considered in our calculations, where the one-body density develops a pronounced multi-peak structure reminiscent of fermionization and the population-transfer fluctuations are strongly suppressed.

\begin{figure}[t]
\centering
\includegraphics[width=7cm]{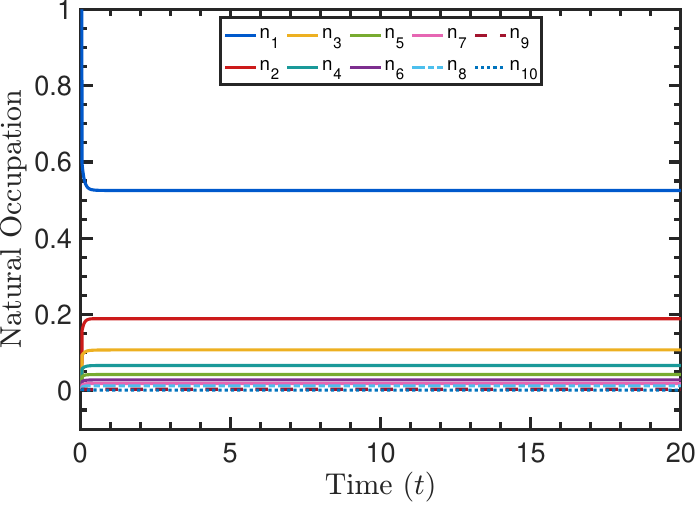}
\caption{Natural-orbital occupations in the strongly interacting regime for $\alpha=3.0$, $g=0.5$, $V_1=0.5$, $V_{\rm imb}=0.2$, and $N=8$. The normalized natural occupations, obtained using $M=10$ time-dependent orbitals, show the redistribution of particles among multiple natural orbitals, demonstrating the development of many-body fragmentation.}
\label{fig11}
\end{figure}

The resulting population-imbalance dynamics are shown in Fig.~\ref{fig10}. Despite the substantial increase in the number of orbitals, $z(t)$ remains remarkably robust. The dynamics exhibit dense, smooth oscillations about a nonzero mean value, and their overall temporal structure remains nearly unchanged as $M$ increases. The most noticeable difference is a shift in the mean value of the population imbalance, whereas the characteristic oscillatory pattern is largely preserved. This indicates that, in the present parameter regime, the population imbalance is relatively insensitive to the inclusion of additional time-dependent orbitals, even though these orbitals enlarge the variational space available to describe the many-body state. In contrast, the natural occupations shown in Fig.~\ref{fig11} reveal that the many-body state has a clear multiorbital character. Several natural orbitals acquire appreciable populations, demonstrating the presence of many-body fragmentation in this strongly interacting regime. The occupation distribution obtained with $M=10$ provides a more extensive description of this multiorbital structure. This fragmentation is consistent with the multi-peak spatial density observed for the same interaction parameters. The robustness of $z(t)$ should therefore not be interpreted as evidence that the underlying many-body dynamics can be adequately described by only a few orbitals. Rather, the natural-orbital occupations demonstrate that several orbitals contribute significantly to the many-body state, even though their inclusion has only a modest effect on the overall temporal structure of the population imbalance. Thus, the population imbalance, as a low-order observable, remains relatively insensitive to the additional many-body degrees of freedom captured by the larger orbital sets, while the natural occupations reveal the strongly correlated and fragmented character of the state.

\begin{figure}[b!]
\centering
\includegraphics[width=8cm]{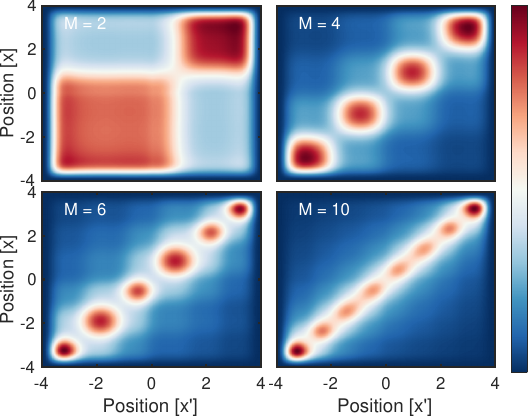}
\caption{One-body reduced density matrix $\rho^{(1)}(x,x';t)$ at $t=20$ for $M=2$, $M=4$, $M=6$, and $M=10$ time-dependent orbitals, calculated for $\alpha=3.0$, $g=0.5$, $V_1=0.5$, $V_{\rm imb}=0.2$, and $N=8$. The increasing orbital number progressively resolves the spatial structure of the one-body density matrix, with eight distinct peaks along the diagonal becoming visible for $M=10$.}
\label{fig:correlation}
\end{figure}

To gain further insight into the orbital-number dependence of the many-body state, we compare the one-body reduced density matrices (1-RDMs) for $M=2$, $4$, $6$, and $10$ at a representative evolution time, $t=20$, chosen arbitrarily for this comparison, as shown in Fig.~\ref{fig:correlation}. The dependence on the orbital number is evident, in marked contrast to the relative robustness of the population-imbalance dynamics. For $M=2$, the 1-RDM is dominated by a broad, bright region, and the multi-peak structure of the initial-state density is not resolved. As $M$ increases to $4$ and $6$, progressively finer spatial features become visible. For $M=10$, eight prominent features are resolved along the diagonal, consistent with the eight-peak structure of the corresponding initial-state density. For each value of $M$, the 1-RDM at $t=20$ remains nearly identical to its initial-state counterpart at $t=0$, consistently with the observed suppression of population-imbalance dynamics.

These observations highlight an important distinction between collective and spatially resolved observables. While the population imbalance retains nearly the same overall temporal structure as $M$ increases, the 1-RDM reveals that a small orbital basis does not adequately resolve the spatial organization of the strongly interacting system. The natural-orbital occupations further demonstrate the significant contribution of multiple orbitals, confirming the multiorbital character of the state. 

\section{Conclusion}\label{sec:conclusion}
In conclusion, we have shown that the nonequilibrium many-body dynamics of a one-dimensional Bose-Josephson junction realized by ultracold atoms in a box potential with a central barrier can be engineered through the interplay of interaction properties, interaction strength, junction coupling, and initial population imbalance. Using the multiconfigurational time-dependent Hartree method for bosons, we have shown that changing the exponent of a finite-core power-law interaction drives a crossover from regular Josephson oscillations to dephasing and recurrent rephasing. This behavior is governed by the full spatial structure of the interaction kernel, rather than simply by its asymptotic range. We further establish that the central barrier and initial population imbalance provide independent means of controlling and suppressing the recurrent dynamics, while increasing the interaction strength can strongly reduce population-transfer fluctuations and promote pronounced many-body spatial structures. The tunability of soft-core interactions, for example through Rydberg dressing, offers a promising route towards realizing interaction-engineered dynamics in ultracold atomic systems.

Comparisons with the two-mode extended Bose-Hubbard model (Appendix \ref{appendix:BH}) further reveal the limits of a fixed two-orbital description. Although it captures the dynamics reasonably well when the many-body evolution is dominated by two modes for the considered barriers, its accuracy deteriorates as stronger interactions activate additional orbital degrees of freedom. At the same time one observes that, for the chosen two-body interaction potential (\ref{eq:interaction}), the two-mode description works better for smaller $\alpha$. The resulting picture highlights the essential role of multiorbital correlations in strongly interacting tunneling dynamics.

As discussed in Appendix \ref{app:contact}, the same sequence of regular oscillations, dephasing, and recurrent rephasing can be observed by varying the strength of a contact interaction. Thus, this dynamical phenomenon is not restricted to a particular interaction profile; it can emerge through distinct routes of interaction engineering. This establishes interaction strength as an alternative control knob and points to a broader robustness of the underlying many-body dynamics.

Several important questions remain open. First, one may investigate to what extent the recurrent rephasing observed in the population imbalance correspond to a genuine recovery of many-body coherence. Although the two interaction-engineering protocols produce remarkably similar population dynamics, they need not generate identical many-body states. Comparing their reduced density matrices, natural-orbital occupations, and many-body fidelities could reveal whether coherence and correlations are also restored during rephasing. It would also be interesting to investigate the quantum dynamics as a function of the power-law decay $\alpha$ at fixed $V_\alpha(0)$, and vice versa, as a function of $V_\alpha(0)$ at fixed long range behavior. Further questions concern the formation of the relative phase between the two subsystems in this one-dimensional BJJ, the accuracy with which it develops, and the extent to which the observed results generalize to larger particle numbers and higher spatial dimensions. Equally important is to establish the robustness of these dynamics against finite temperature, particle loss, and imperfections in the trapping potential. Addressing these questions would clarify the microscopic origin of recurrent rephasing and assess its feasibility as a controllable feature of nonequilibrium dynamics in experimentally accessible ultracold atomic systems.

\section*{Acknowledgments}
A.~K.~S. acknowledges support from JSPS KAKENHI Grant Number JP25K00925.

\appendix

\section{MCTDH-X units}\label{app:units}

We introduce a characteristic length $\ell$ and express the Schrödinger equation in dimensionless form. The Hamiltonian is rescaled by the characteristic energy $E_0=\hbar^2/m\ell^2$, which defines the corresponding natural time scale as $t_0=\hbar/E_0=m\ell^2/\hbar$. Accordingly, all lengths, energies, and times in the numerical calculations are expressed in units of $\ell$, $E_0$, and $t_0$, respectively. To connect the dimensionless quantities with experimentally relevant scales, we consider $^{87}$Rb atoms and choose $\ell=10~\mu\mathrm{m}$, consistent with characteristic length scales encountered in ultracold-atom experiments \cite{Gaunt2013,Albeiz:2005,Navon2015}. For this choice, the characteristic energy and time scales are $E_0\simeq 7.7\times10^{-34}$~J and $t_0\simeq 0.137$~s, respectively. The corresponding frequency scale is $E_0/h\simeq 1.16$~Hz. The dimensionless parameters and their ranges used in the calculations are summarized in Table~\ref{tab:dimensionless_parameters}.

\begin{table}[htbp]
\centering
\caption{Dimensionless parameters and their ranges used in the MCTDHB calculations.}
\label{tab:dimensionless_parameters}
\begin{tabular}{lcl}
\hline\hline
Parameter & Range/value & Description\\
\hline
$L$ & $4$ & Half-length of the box\\
$\sigma$ & $0.1$ & Gaussian barrier width\\
$V_0$ & $60$ & Outer confining barrier height\\
$V_1$ & $0.5$--$3.0$ & Central barrier height\\
$V_{\rm imb}$ & $0.2$--$2.0$ & Initial imbalance-potential amplitude\\
$\alpha$ & $0.5$--$4.0$ & Interaction exponent\\
$g$ & $0.01$--$0.5$ & Interaction strength\\
$d^3$ & $0.05$ & Finite-core parameter\\
$d$ & $\simeq 0.368$ & Interaction-core length\\
$N$ & $8$, $10$ & Number of bosons\\
$M$ & $4$--$10$ & Number of time-dependent orbitals\\
\hline\hline
\end{tabular}
\end{table}

\section{Comparison with the two-mode extended Bose-Hubbard model}\label{appendix:BH}

To assess the necessity of a multi-orbital description for accurately describing the continuum many-body dynamics, we compare compare some of the MCTDHB results presented in Sec.~\ref{sec:results} with those obtained from a two-mode extended Bose-Hubbard model. This provides a simplified lattice description in which the dynamics is expressed in terms of tunneling between localized modes and interaction-induced on-site energy shifts. Its applicability is expected to depend on whether the dynamics remain predominantly within the lowest single-particle orbital of each well.

We consider the same trapping geometry, interaction parameters, and initial-state preparation as in the continuum calculations. The comparison focuses on the time evolution of the population imbalance for different interaction strengths. This allows us to identify the parameter regime in which the reduced lattice description reproduces the continuum dynamics and the regime in which the contribution of higher orbitals becomes important.

The Hamiltonian we consider is \cite{Li, Pizzardo, Vianello2026}
\begin{align}
\hat{H}_{\mathrm{BH}} &= -J\left(\hat{a}_L^\dagger\hat{a}_R+ \hat{a}_R^\dagger\hat{a}_L\right) + \frac{U}{2}\sum_{\ell=L,R} \hat{n}_\ell(\hat{n}_\ell-1) \nonumber\\
&+ V \hat n_L \hat n_R + T [\hat a_L^\dag (\hat n_L+\hat n_R) \hat a_R + {\rm H.c.}] + P (\hat a_L^\dag \hat a_L^\dag \hat a_R \hat a_R + {\rm H.c.}),
\label{eq:BH}
\end{align}
where $\hat{a}_\ell^\dagger$ and $\hat{a}_\ell$ are the bosonic creation and annihilation operators in the left ($L$) and right ($R$) wells, respectively, $\hat{n}_\ell=\hat{a}_\ell^\dagger\hat{a}_\ell$ is the corresponding number operator, and H.c.~denotes the Hermitian conjugate. The total particle number is conserved, with $\hat{n}_L+\hat{n}_R=N$. The parameters $J$ and $U$ represent the single-particle tunneling amplitude and the on-site inter-action energy, respectively. The former originates from the one-body term of the many-body Hamiltonian, while the two-body term is responsible for the on-site interaction $U$, the nearest-neighbor density-density interaction $V$, the collisionally induced tunneling $T$, and the pair tunneling $P$.

The localized left- and right-well orbitals are constructed from the even-symmetry ground state $\phi_g(x)$ and the odd-symmetry first excited state $\phi_u(x)$ of the single-particle Hamiltonian $h(x)$ as
\begin{equation}
\phi_{L,R}(x)=\frac{1}{\sqrt{2}}
\left[\phi_g(x)\pm\phi_u(x)\right].
\end{equation}
The tunneling amplitude is obtained from the matrix element
\begin{equation}
J=-\int dx\,\phi_L^*(x)h(x)\phi_R(x).
\label{eq:J_parameter}
\end{equation}
For a symmetric double-well potential, this expression is equivalently written as
\begin{equation}
J=\frac{\epsilon_u-\epsilon_g}{2},
\label{eq:J_energy}
\end{equation}
where $\epsilon_g$ and $\epsilon_u$ are the single-particle energies associated with $\phi_g$ and $\phi_u$, respectively.

\begin{figure}[t]
\centering
\includegraphics[width=11cm]{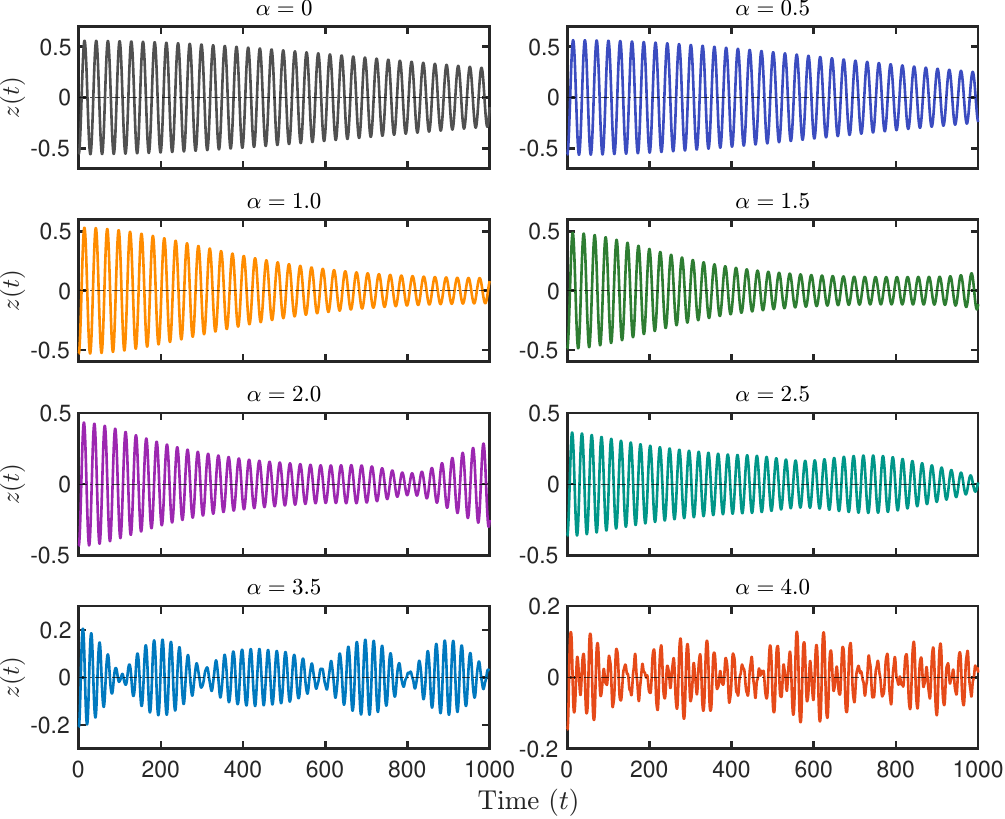}
\caption{Population-imbalance dynamics $z(t)$ obtained from the eBH model for different interaction exponents $\alpha$. The parameters are $g=0.01$, $V_1=0.5$, and $V_{\rm imb}=0.2$, with $N=10$ bosons.}
\label{fig:BH_comparison_diff_alpha_z}
\end{figure}

The other parameters are calculated by projecting the finite-core power-law interaction onto the localized orbitals:
\begin{align}
U &= \int dx\,dx'\,|\phi_L(x)|^2 V_\alpha(x-x')|\phi_L(x')|^2,\\
T &= \int dx\,dx'\, \phi_L(x) \phi_R(x) V_\alpha(x-x') |\phi_L(x')|^2, \\
P &= \frac{1}{2} \int dx\,dx'\, \phi_L(x) \phi_R(x) V_\alpha(x-x') \phi_L(x') \phi_R(x'), \\
V &= \int dx\,dx'\, |\phi_L(x)|^2 V_\alpha(x-x') |\phi_R(x')|^2 + 2P,
\end{align}
where
\begin{equation}
V_\alpha(x-x')=\frac{g}{|x-x'|^\alpha+d^\alpha}.
\end{equation}

The population imbalance is defined as
\begin{equation}
z(t)=\frac{\langle\hat{n}_L(t)\rangle-
\langle\hat{n}_R(t)\rangle}{N}.
\label{eq:BH_imbalance}
\end{equation}

We compare the population-imbalance dynamics obtained from the eBH model with the continuum MCTDHB results in two representative interaction regimes. The first corresponds to weak interactions, where the dynamics exhibit regular Josephson oscillations followed by interaction-induced dephasing and recurrent rephasing. The second corresponds to the strongly interacting regime, where population-transfer fluctuations are strongly suppressed and the many-body state exhibits pronounced multiorbital character.

\begin{figure}[t]
\centering
\includegraphics[width=7cm]{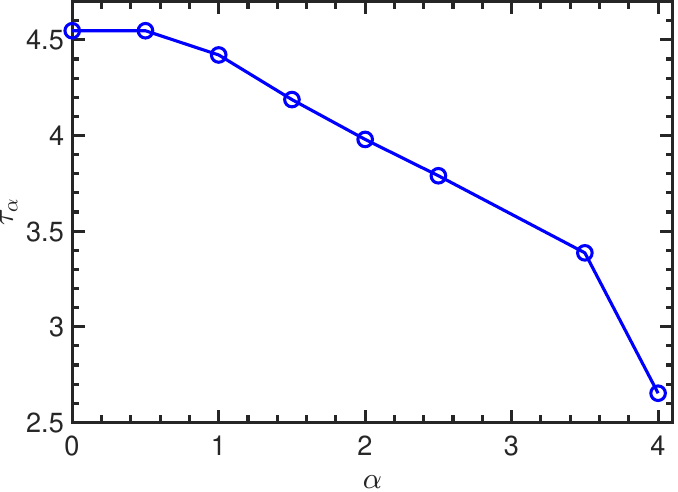}
\caption{Characteristic timescale $\tau=2\pi/\omega$, extracted from the dominant frequency of the population-imbalance dynamics obtained from the eBH model, as a function of the interaction exponent $\alpha$. The parameters are $g=0.01$, $V_1=0.5$, and $V_{\rm imb}=0.2$, with $N=10$ bosons.}
\label{fig:BH_comparison_tau}
\end{figure}

We first consider the weakly interacting regime, where the dynamics exhibit a crossover from regular Josephson oscillations to dephasing and subsequent recurrent rephasing as the interaction exponent $\alpha$ increases. Fig.~\ref{fig:BH_comparison_diff_alpha_z} reports the eBH dynamics of $z(t)$, which should be compared with the corresponding MCTDHB dynamics shown in Fig.~\ref{fig2}. The eBH description reproduces qualitatively the different stages of the dynamics, capturing the transition from regular oscillations to dephasing and the subsequent rephasing, although the quantitative agreement degrades at larger $\alpha$. The corresponding characteristic timescale $\tau=2\pi/\omega$, extracted from the dominant frequency of $z(t)$, is shown in Fig.~\ref{fig:BH_comparison_tau} and is in close agreement with the MCTDHB result reported in Fig.~\ref{fig3}(d). This qualitative agreement can be understood from the natural-orbital occupations obtained in the MCTDHB calculations for this weakly interacting regime, where the two lowest orbitals contribute predominantly to the many-body state, while the occupations of higher orbitals remain negligible. Consequently, the dynamics are effectively confined to a two-mode space, providing a physical basis for the success of the eBH description.

\begin{figure}[t]
\centering
\includegraphics[width=11cm]{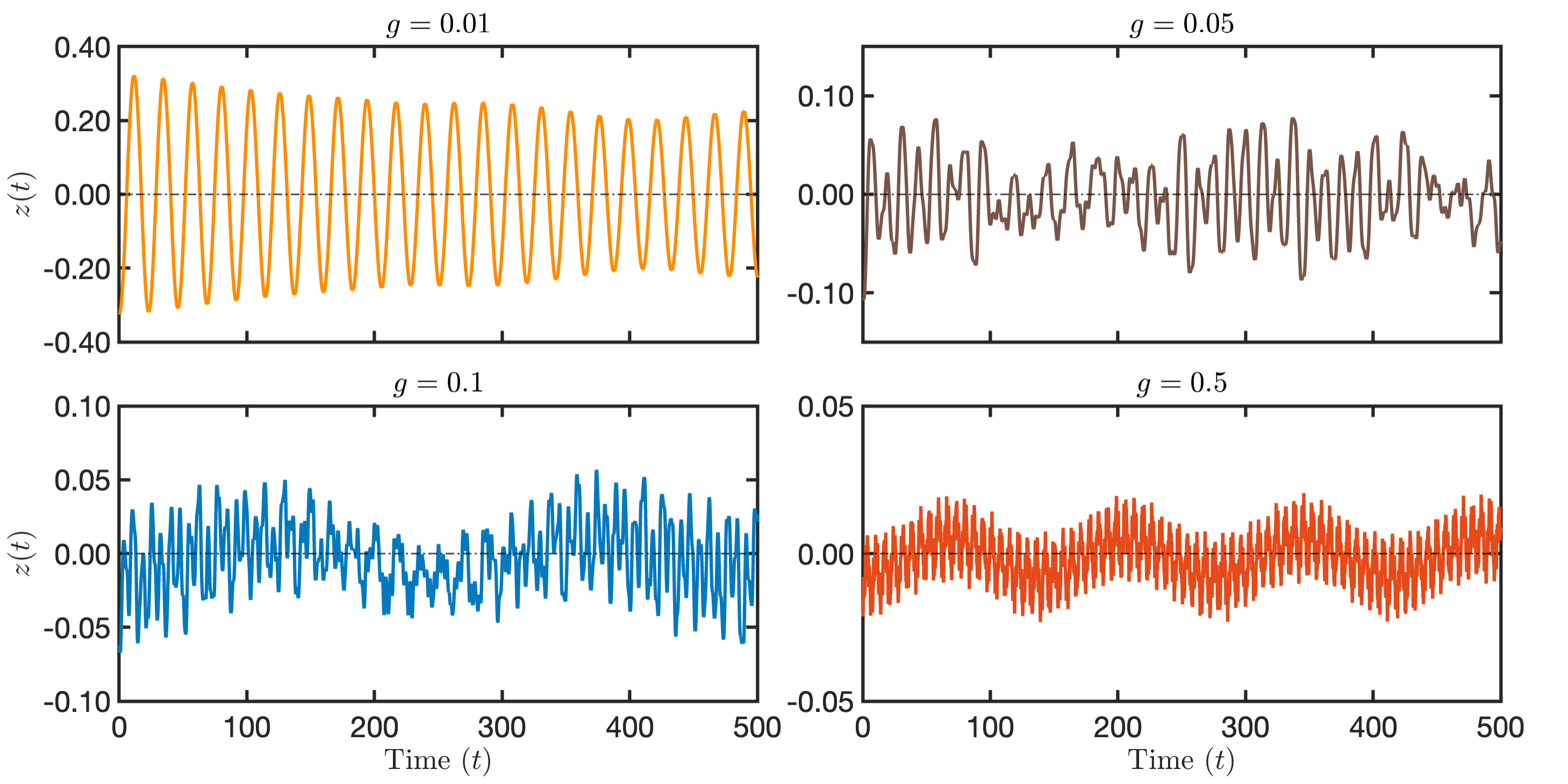}
\caption{Population-imbalance dynamics $z(t)$ obtained from the eBH model for increasing interaction strength $g=0.01$--$0.5$, at fixed interaction exponent $\alpha=3.0$. The remaining parameters are $V_1=0.5$, $V_{\rm imb}=0.2$, and $N=8$.}
\label{fig:BH_comparison_alpha_3}
\end{figure}

We next examine how the agreement between the eBH and MCTDHB descriptions changes with increasing interaction strength at fixed $\alpha=3.0$. Fig.~\ref{fig:BH_comparison_alpha_3} shows the population-imbalance dynamics obtained from the eBH model for interaction strengths ranging from $g=0.01$ to $g=0.5$, while the corresponding MCTDHB results are presented in Fig.~\ref{fig8}. At the weakest interaction, $g=0.01$, the two descriptions are still in qualitative agreement, indicating that the dynamics can be captured within an effective two-mode space. As $g$ increases, however, the eBH results neatly depart from the MCTDHB dynamics. In the strongly interacting regime, MCTDHB captures the population dynamics together with the increasingly structured, fermionization-like density profile shown in Fig.~\ref{fig7}, whereas the eBH description becomes inadequate. This departure reflects the increasing importance of multiorbital contributions to the many-body dynamics. At weak coupling, the dynamics are predominantly supported by the two leading natural orbitals, making a two-mode description effective. With increasing interaction strength, additional orbitals become relevant, allowing MCTDHB to describe the spatial rearrangement and correlations associated with the fermionized-like regime.

\begin{figure}[ht!]
\centering
\includegraphics[width=11cm]{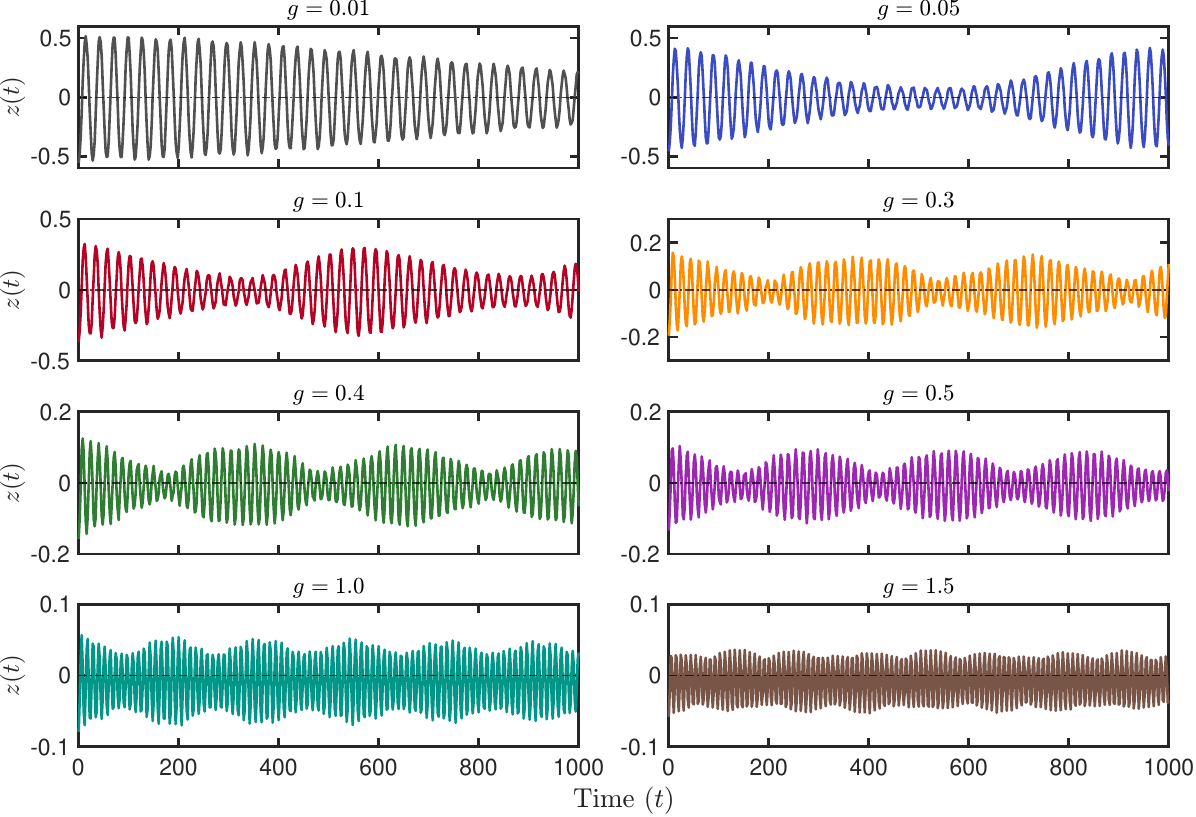}
\caption{Time evolution of the population imbalance $z(t)$ for different strengths $g$ of the contact interaction, $W(x-x')=g\delta(x-x')$. The trapping potential and initial-state preparation are identical to those used in the finite-core power-law interaction calculations. }
\label{appendix:contact_z(t)}
\end{figure}

\begin{figure}[ht!]
\centering
\includegraphics[width=11cm]{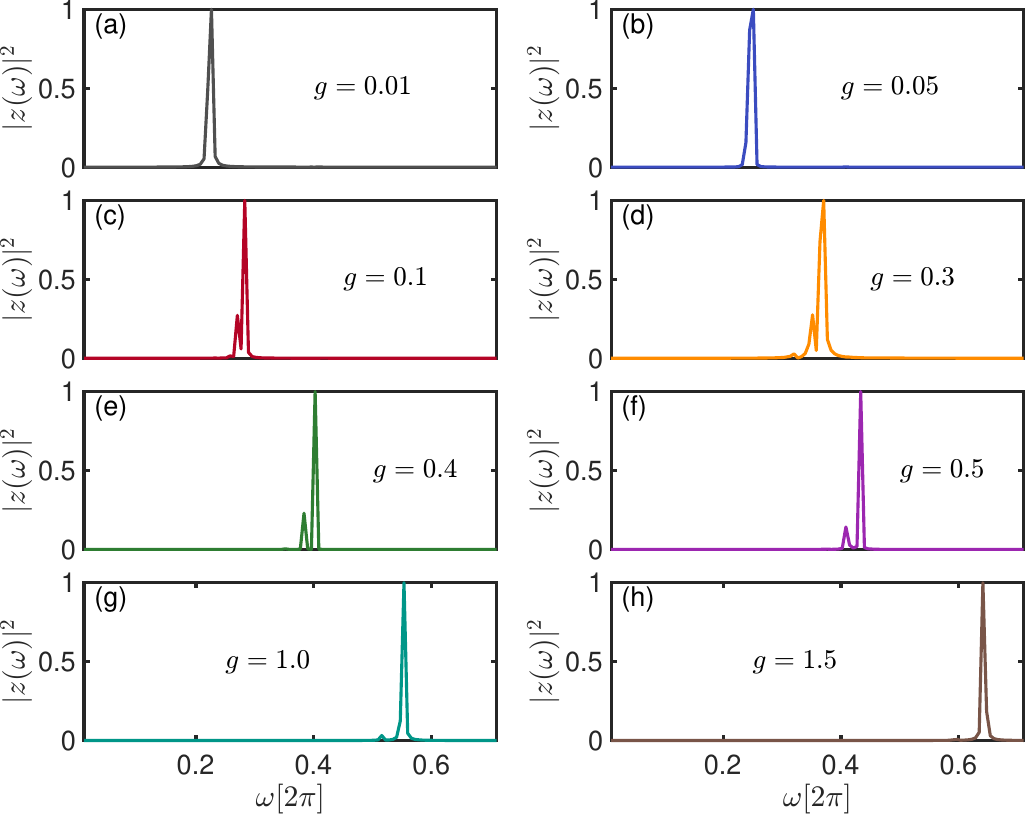}
\caption{Frequency spectra of the population-imbalance dynamics for different contact-interaction strengths $g$. The spectra illustrate the evolution of the dominant frequency components across the regular, dephasing, and recurrent-rephasing regimes. }
\label{appendix:contact_frequency}
\end{figure}

\section{Comparison with contact interaction}\label{app:contact}

\begin{figure}[t]
\centering
\includegraphics[width=7cm]{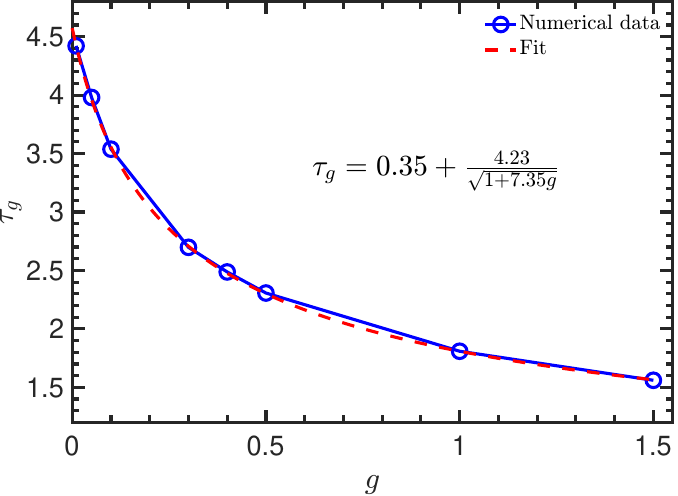}
\caption{Characteristic timescale $\tau=2\pi/\omega$ as a function of the contact-interaction strength $g$, where $\omega$ is the dominant frequency extracted from the corresponding population-imbalance spectrum. The numerical data are well described by an inverse-square-root fit $\tau_g = 0.35 + 4.23/\sqrt{1+7.35 g}$.}
\label{appendix:contact_tau}
\end{figure}

Here we briefly compare some of the results presented in Sec.~\ref{sec:results} with the dynamical behavior obtained by tuning the interaction strength alone \cite{Saha2026}. To address this question, we retain the trapping potential and initial-state preparation used above, but replace the finite-core power-law interaction with a contact interaction, $W(x-x')=g\delta(x-x')$, and vary its strength $g$. One sees that by increasing $g$ a similar sequence of dynamical regimes is observed as when varying $\alpha$ in \eqref{eq:interaction}, namely regular Josephson oscillations, followed by dephasing and the subsequent emergence of recurrent rephasing. This behavior is seen in the population-imbalance dynamics shown in Fig.~\ref{appendix:contact_z(t)}. The corresponding frequency spectra in Fig.~\ref{appendix:contact_frequency} reveal how the spectral content evolves across these regimes. Consistently, the characteristic timescale $\tau=2\pi/\omega$, extracted from the dominant frequency $\omega$, exhibits a systematic dependence on $g$, as shown in Fig.~\ref{appendix:contact_tau}. We emphasize the quantitative difference in the functional dependence of $\tau_g$ on $g$ with respect to the functional dependence of $\tau_\alpha$ on $\alpha$. A systematic study of to what extent one can reproduce many-body tunneling dynamics in presence of long-range interaction and finite-core with a suitably chosen contact interaction is an interesting question to be investigated.

\bibliographystyle{iopart-num}
\bibliography{biblio}

\end{document}